\pdfoutput=1
\newcommand*{\ATLASLATEXPATH}{}
\documentclass[PAPER, cernpreprint, atlasdraft=false, texlive=2016, UKenglish]{\ATLASLATEXPATH atlasdoc}
 
\usepackage{\ATLASLATEXPATH atlaspackage}
\usepackage{\ATLASLATEXPATH atlasbiblatex}
 
\usepackage{\ATLASLATEXPATH atlasphysics}
 
\graphicspath{{logos/}{figures/}}
 
\usepackage{ANA-EXOT-2018-15-PAPER-defs}
\usepackage{multirow}
\usepackage{tabularx,amsmath,stackengine}
\usepackage[makeroom]{cancel}

\AtlasTitle{Search for a right-handed gauge boson decaying into a high-momentum
heavy neutrino and a charged lepton in $pp$ collisions with the ATLAS detector at $\sqrt{s}=13$~\TeV}

\AtlasJournalRef{Phys. Lett. B 798 (2019) 134942}
\AtlasDOI{10.1016/j.physletb.2019.134942}

\AtlasAbstract{
A search for a right-handed gauge boson \WR, decaying into a
boosted right-handed heavy neutrino \NR,
in the framework of Left-Right Symmetric Models is presented.
It is based on
data from proton--proton collisions
with a centre-of-mass energy of $13$~\TeV\
collected by the ATLAS detector at the Large Hadron Collider
during the years 2015, 2016 and 2017, corresponding to an integrated
luminosity of $80$~fb$^{-1}$.
The search is performed separately for electrons and muons in the final state.
A distinguishing feature of the search
is the use of large-radius jets containing electrons.
Selections based on the signal topology result in smaller background compared to the expected signal.
No significant deviation from the Standard Model prediction is observed and
lower limits are set in the \WR and \NR mass plane.
Mass values of the \WR smaller than 3.8--5~\TeV\ are excluded for \NR in the mass range 0.1--1.8~\TeV.
}
 
\author{The ATLAS Collaboration}
 
\AtlasRefCode{EXOT-2018-15}

\PreprintIdNumber{CERN-EP-2019-052}

\AtlasCoverSupportingNote{Search for a boosted right-handed heavy neutrino with the ATLAS detector}{https://cds.cern.ch/record/2266957}
 
\AtlasCoverCommentsDeadline{27 March 2019}
 
\AtlasCoverAnalysisTeam{Deepak Kar, Debarati Roy, Xifeng Ruan}
 
\AtlasCoverEdBoardMember{Sarah Heim (chair), Pierre-Antoine Delsart, Leonid Serkin}

\AtlasCoverEgroupEditors{atlas-EXOT-2018-15-editors@cern.ch}
 
\AtlasCoverEgroupEdBoard{atlas-EXOT-2018-15-editorial-board@cern.ch}

\hypersetup{pdftitle={Search for a right-handed gauge boson decaying into a high-momentum
heavy neutrino and a charged lepton in $pp$ collisions with the ATLAS detector at $\sqrt{s}=13$~\TeV},pdfauthor={The ATLAS Collaboration}}

\begin{document}
 
\maketitle

\section{Introduction}
\label{sec:intro}

Over the past decades, there have been several important developments at the theoretical and
experimental frontiers to address the question of neutrino mass generation, which is not explained in
the Standard Model (SM) of particle interactions.
A widely adopted approach to explain small neutrino masses is the so-called
seesaw mechanism \cite{Weinberg:1979sa},
where the light neutrinos acquire their Majorana masses from
dimension-5 operators through electroweak symmetry breaking.
The simplest seesaw mechanism can be categorised into a few different
classes, such as the Type-I \cite{Minkowski:1977sc, Yanagida:1979as, GellMann:1980vs},
Type-II \cite{Magg:1980ut,PhysRevD.22.2860,PhysRevD.23.165}
and Type-III \cite{Magg:1980ut, Foot:1988aq} seesaw scenarios. Type-I and Type-II models can further be
embedded into a Left-Right Symmetric Model
(LRSM) \cite{Pati:1974yy, Mohapatra:1974gc, Senjanovic:1975rk}.
The LRSM contains SM-singlet heavy neutrinos \NR, which are introduced as the parity gauge partners
of the corresponding left-handed neutrino fields, and a right-handed gauge boson \WR.

\begin{figure}[h]
\centering
\includegraphics[width=0.4\textwidth]{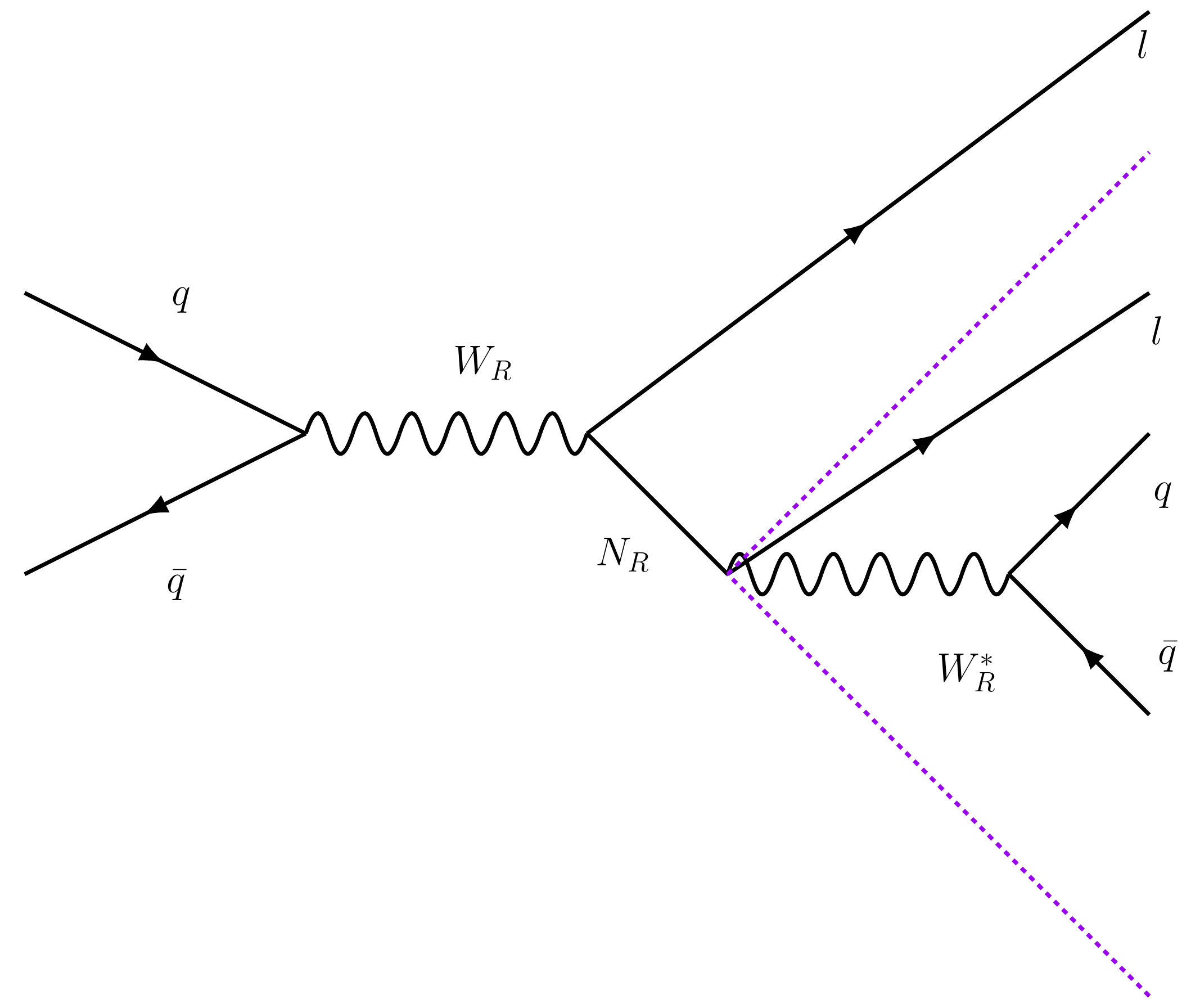} \qquad
\caption{Diagram of the \WR decay via \NR into charged leptons and quarks.
The leptons need to be of the same flavour, but can be the
same or opposite charges.
The dashed purple lines indicate that the \NR decay products
can be inside a \LRJ.}
\label{fig:fd}
\end{figure}

The LRSM framework provides a natural set-up for the seesaw mechanism and offers
several features including
parity symmetry at high energy, mass generation of the light and heavy neutrinos,
explanation of parity violation in the SM and existence of the right-handed charged current.
This model can naturally explain the small neutrino masses through the Type-I seesaw via the right-handed neutrinos,
and the Type-II seesaw via SU(2)-triplet scalars. Both the Type-I and Type-II contributions can coexist.
In the LRSM, left-handed neutrinos (SM neutrinos) as well as the right-handed
neutrinos are considered to be Majorana particles (i.e.\ to be their own antiparticles).
The LRSM thus features violation of the global lepton number symmetry of the SM.
Hence, the model can be tested by observing lepton-number-violating processes,
such as the Keung--Senjanovi\'{c} process~\cite{PhysRevLett.50.1427}, shown in Figure~\ref{fig:fd}.
 
Searches by the ATLAS~\cite{Aad:2015xaa, Aaboud:2018spl} and
CMS~\cite{Khachatryan:2014dka, Khachatryan:2016jqo, Sirunyan:2018pom, Sirunyan:2018xiv, Sirunyan:2018vhk} collaborations
for signatures of LRSMs
have considered the final state containing two charged leptons and two jets and
have excluded regions of the (\MWR,\,\MNR) parameter space for \MWR and \MNR up to several
\TeV, where \MNR and \MWR denote
the masses of \NR and \WR, respectively.

This search is focused on the regime where the \WR is very
heavy compared with the \NR ($\MNR/\MWR \le 0.1$),
and investigates an alternative signature for $\WR \rightarrow \NR \ell$ decays, following Ref.~\cite{Mitra:2016kov}.
The probed mass regime enables exploration of a parameter space complementary to the one used in
previous searches that reconstruct the \NR decay into a charged lepton and two jets,
later referred to as the ``resolved topology''.
In the probed mass regime, the heavy neutrinos are produced with large transverse momentum (i.e. are highly boosted)
and their decay products are very collimated.
Therefore a large-radius jet (\LRJ) can be used to reconstruct all or part of the \NR.
Since jet construction in ATLAS
includes the energy deposition of electrons in the
calorimeters but no muons, the analysis strategy
is different for the two cases. In the electron channel, the electron
energy deposit is included in the constructed \LRJ
originating
from the decay of the \NR, and the \LRJ can be considered as a proxy for the \NR. In the muon channel, the four-momentum
of the muon is added to the \LRJ to obtain the \NR four-momentum.
The search is restricted to the scenarios where both leptons have the same flavour.
No constraint on their charge is enforced, because of the higher probability of
charge misidentification for high-\pT electrons.

The results obtained in this search are also applicable to other
variations of the LRSM that contain a right-handed gauge boson and neutral leptons,
such as inverse seesaw models~\cite{Dias:2012xp}.
Additionally, this search is also applicable to R-parity-violating supersymmetry~\cite{Barbier:2004ez, Aaboud:2017faq},
where a selectron
is resonantly produced and subsequently decays into an
electron and a neutralino, and the
latter decays to a lepton and quarks through a non-zero $\lambda^{\prime}$ coupling.
When the neutralino is boosted, its decay products can be reconstructed as a single \LRJ~\cite{Allanach:2014nna}, analogous to
the final state probed in this analysis.

\section{ATLAS detector}
\label{sec:atlassdet}
 
The ATLAS detector~\cite{:2008zzm} at the Large Hadron Collider (LHC)
is a multipurpose particle detector with a forward--backward symmetric
cylindrical geometry and a near $4\pi$ coverage in solid angle.\footnote{
ATLAS uses a right-handed coordinate system with its origin at the nominal interaction point (IP)
in the centre of the detector and the $z$-axis along the beam pipe.
The $x$-axis points from the IP to the centre of the LHC ring, and the $y$-axis points upwards.
Cylindrical coordinates $(r,\phi)$ are used in the transverse plane, $\phi$ being the azimuthal angle around the $z$-axis.
The pseudorapidity is defined in terms of the polar angle $\theta$ as $\eta = -\ln \tan(\theta/2)$.
The angular separation between two objects is defined as
$\Delta R \equiv \sqrt{(\Delta \eta)^{2} + (\Delta\phi)^{2}}$,
where $\Delta \eta$ and $\Delta\phi$ are the separations in $\eta$ and $\phi$.
The rapidity is defined as $y = \frac{1}{2} \ln \frac{E + p_{\mathrm{z}}}{E - p_{\mathrm{z}}}$, where $E$ is the energy
and $p_{\mathrm{z}}$ is the longitudinal component of the momentum along the beam
pipe.
The angular separation between two objects in terms of rapidity is defined as
$\Delta R_{\mathrm{y}} \equiv \sqrt{(\Delta y)^{2} + (\Delta\phi)^{2}}$,
where $\Delta y$ and $\Delta\phi$ are the separations in $y$ and $\phi$.
Momentum in the transverse plane is denoted by \pT.}
It consists of an inner tracking detector (ID) surrounded by a thin superconducting solenoid providing a $2$ T axial
magnetic field, electromagnetic (EM) and hadronic calorimeters, and a muon spectrometer (MS).
The ID consists of silicon pixel, silicon microstrip, and straw-tube transition-radiation tracking detectors, covering
the pseudorapidity range $|\eta| < 2.5$.
The calorimeter system covers the pseudorapidity range $|\eta| < 4.9$.
Electromagnetic calorimetry is provided by barrel and endcap
high-granularity lead and liquid-argon (LAr) sampling calorimeters, within the region $|\eta| < 3.2$.
There is an additional thin LAr presampler covering $|\eta| < 1.8$, to correct for
energy loss in material upstream of the calorimeters.
For $|\eta| < 2.5$, the LAr calorimeters are divided into three layers in depth.
Hadronic calorimetry is provided by a steel/scintillator-tile calorimeter,
segmented into three barrel structures within $|\eta| < 1.7$, and two copper/LAr hadronic endcap calorimeters,
which cover the region $1.5 < |\eta| < 3.2$.
The forward solid angle up to $|\eta| = 4.9$ is covered by copper/LAr and tungsten/LAr calorimeter
modules, which are optimised for energy measurements of electrons/photons and hadrons, respectively.
The muon spectrometer is the outermost layer of the detector, and is designed to measure
muons up to $|\eta|$  of 2.7. It
comprises separate trigger and high-precision tracking chambers
that measure the deflection of muons in a magnetic field generated by superconducting air-core toroids.
The muon trigger chambers cover up to $|\eta|$  of 2.4.
 
The ATLAS detector selects events using a tiered trigger system~\cite{Aaboud:2016leb}.
The first level is implemented in custom electronics and reduces the event rate
from the bunch-crossing frequency
of 40~MHz to a design value of 100~kHz.
The second level is implemented in software, running on a general-purpose processor
farm which processes the events and reduces the rate of recorded events to about 1~kHz.
 
\section{Data and simulation samples}
\label{sec:mc}
 
This analysis uses proton--proton ($pp$) collision data at a centre-of-mass energy $\sqrt{s} =13$~\TeV\
collected in 2015, 2016 and 2017
that satisfy a number of data-quality criteria.
The amount of data used in this analysis corresponds to an integrated luminosity of $80$~fb$^{-1}$.
 
Simulated signal and background events are used to optimise the event selection, to validate
the performance of \LRJS containing an electron, evaluate
the $Z$+jets
background contribution, and calculate signal yields and their systematic uncertainties.
Signal events were simulated at leading order (LO) in QCD using
\amcnlo
2.2.2~\cite{Alwall:2014hca},
with \pythia 8.186~\cite{Sjostrand:2007gs}
using the NNPDF23LO~\cite{Ball:2012cx} parton distribution function (PDF) set and the A14 set of
tuned parameters (tune)~\cite{:a14}
for parton showering and hadronisation.
A version of a LRSM model produced with FeynRules~\cite{Alloul:2013bka} was
implemented~\cite{Roitgrund:2014zka} in \amcnlo and further modified by the authors
of Refs.~\cite{Maiezza:2015lza, Nemevsek:2016enw}.
This model assumes the equivalence of left and right-handed weak gauge couplings,
universality of all the right-handed quarks and right-handed leptons, and the same
masses for all three flavours of heavy right-handed neutrinos.
Events were generated without constraints on the charge of leptons,
in line with the production of Majorana neutrinos.
Signal samples were generated for different \MWR and \MNR hypotheses,
covering the range of 3--6~\TeV\ for \MWR and 150--600~\GeV\ for \MNR.
The production cross-sections are scaled to next-to-leading order (NLO) in QCD following Ref.~\cite{Mattelaer:2016ynf}.

The background processes considered are top-quark pairs ($t\overline{t}$), $Z/W$+jets, single top-quark, diboson and multijet production.
Table~\ref{tab:bgmc} summarises the generator configurations used to produce the samples.
The $t\overline{t}$ sample cross-sections are scaled to next-to-next-to-leading order (NNLO)
in perturbative QCD, including soft-gluon resummation to next-to-next-to-leading-log (NNLL) accuracy~\cite{Czakon:2011xx},
assuming a top-quark mass $m_{t} =172.5$~\GeV~\cite{topmc}.
The resummation damping parameter, $h_{\textrm{damp}}$ in the \powheg model,
which controls the matching of matrix elements to parton showers and regulates
the high-\pT radiation, was set to $1.5 m_{t}$.
The single-top-quark and $W/Z+$jets samples are scaled to the NNLO
theoretical cross-sections~\cite{Kidonakis:2011wy,Kidonakis:2010ux,Kidonakis:2010tc,Anastasiou:2003ds}.
 
\begin{table*}[ht]
\footnotesize
\begin{center}
\caption{Main features of the Monte Carlo models used to simulate background samples.
Top quark refers to both the $t\overline{t}$  and single-top processes.
ME and PS refer
to matrix element and parton shower, while leading order (LO) and next-to-leading order (NLO) indicate the
accuracy of the generators in perturbative QCD (pQCD).
For \powheg+\pythiaeight, different PDF sets were used in ME and PS.}
\begin{tabular}{l | c | c | l l | c}
\toprule
Process & Top quark &  $W$+jets & $Z$+jets &  Diboson &  Multijet  \\
\midrule
Generator & \powheg~\cite{Nason:2004rx, Frixione:2007nw, Frixione:2007vw, Alioli:2010xd}+\pythiaeight &  \powheg+\pythiaeight &  \multicolumn{2}{|c|}{\sherpa~\cite{Gleisberg:2008ta}} &  \pythiaeight \\
ME order in pQCD & NLO & NLO &  \multicolumn{2}{|c|}{NLO}  & LO \\
Version &  v2, 8.186 &   v2, 8.186   &  \multicolumn{2}{|c|}{2.2.1} & 8.186\\
PDF (ME, PS) & NNPDF30NLO~\cite{Ball:2014uwa}, NNPDF23LO &  CT10~\cite{Lai:2010vv}, CTEQ6L1 &\multicolumn{2}{|c|}{NNPDF30NNLO} & NNPDF23LO \\
PS tune & A14 & AZNLO~\cite{Aad:2014xaa} & \multicolumn{2}{|c|}{Default} & A14 \\
\bottomrule
\end{tabular}
\label{tab:bgmc}
\end{center}
\end{table*}
 
The MC samples were processed through the full ATLAS detector simulation ~\cite{Aad:2010ah}
based on \geant~\cite{Agostinelli:2002hh},
or a faster simulation~\cite{ATLAS:1300517} based on a parameterisation
of the calorimeter response and \geant for the other detector systems,
and reconstructed and analysed using the same procedure and software as used for the data.
The signal modelling is found to be consistent between the full and the fast simulation, after
application of dedicated calibration procedures.
To simulate the effects of additional collisions in the same and neighbouring bunch crossings (pile-up),
additional minimum-bias events generated using \pythiaeight with the A3 tune~\cite{ATL-PHYS-PUB-2016-017}
and MSTW2008~\cite{Martin:2009iq} PDF
set were overlaid onto the signal and background simulated events, with a distribution of the number of collisions
matching that of the data.
To account for the differences in particle reconstruction, trigger, identification and isolation efficiencies between simulation and data,
correction factors are derived in dedicated measurements and applied to the simulated events.

\section{Event selection and characterisation}
\label{sec:sel}

\begin{figure}[h!]
\centering
\includegraphics[width=0.37\textwidth]{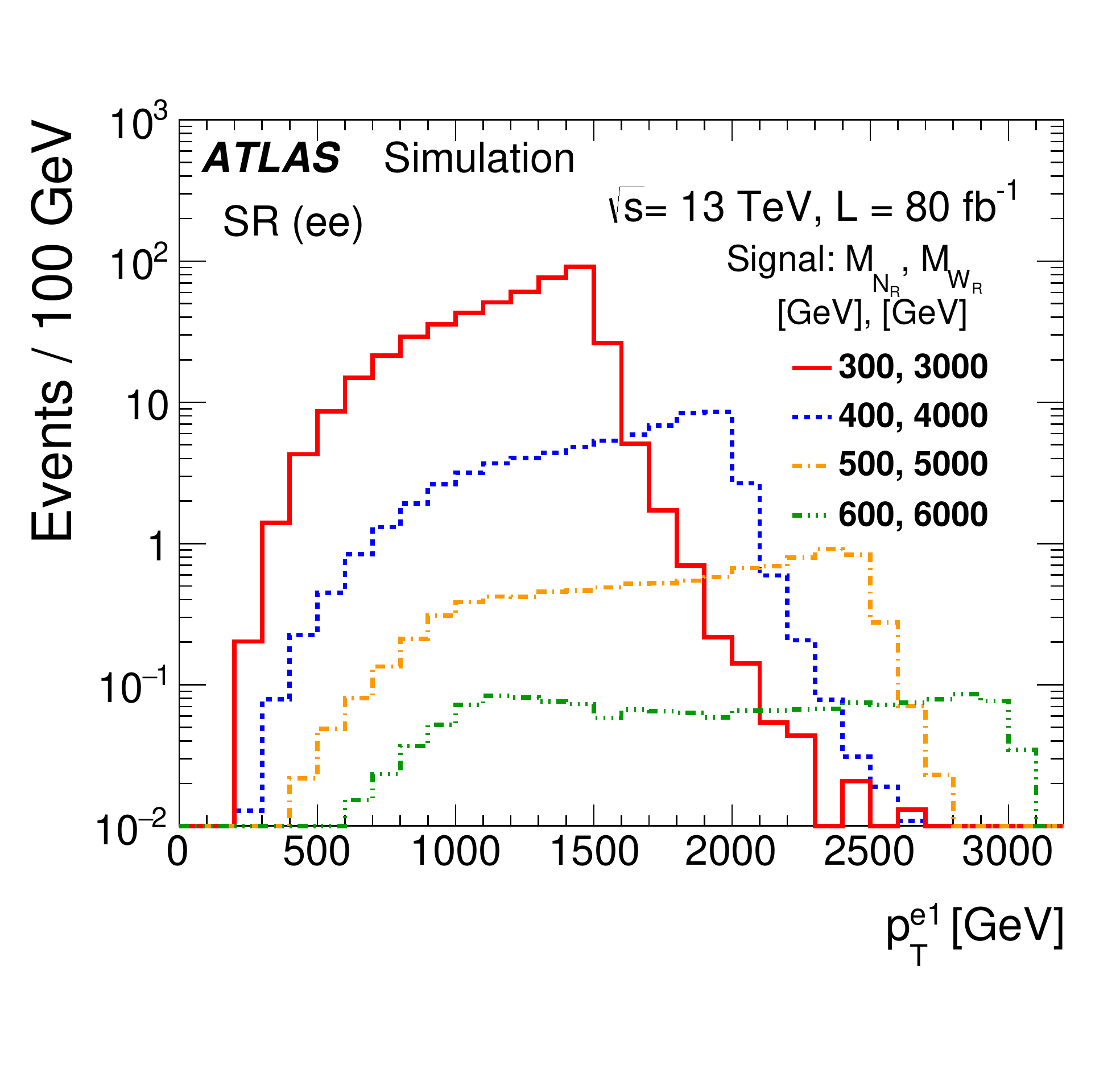} \qquad
\includegraphics[width=0.37\textwidth]{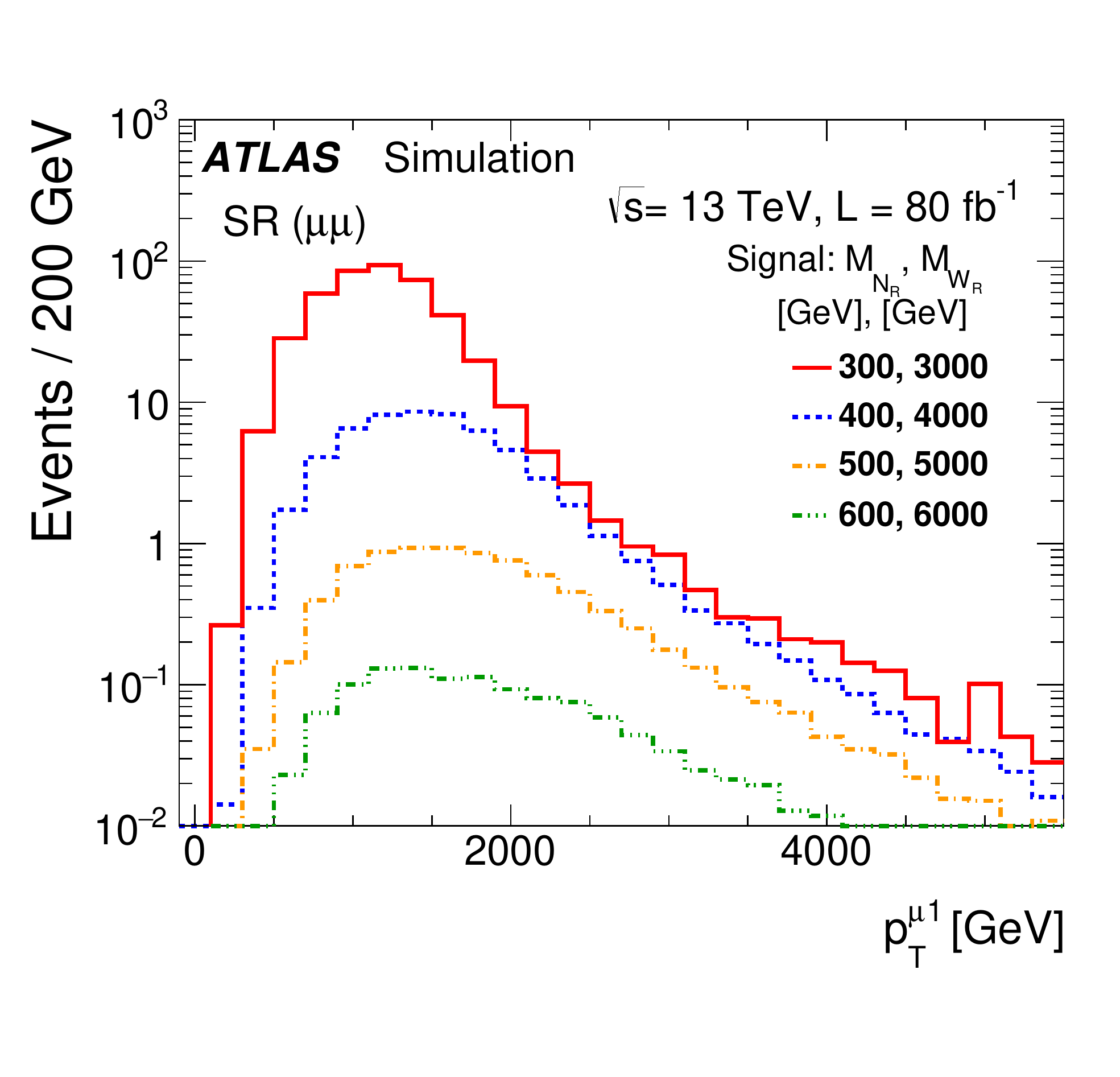} \\[1em]
\vspace{-1cm}
\includegraphics[width=0.37\textwidth]{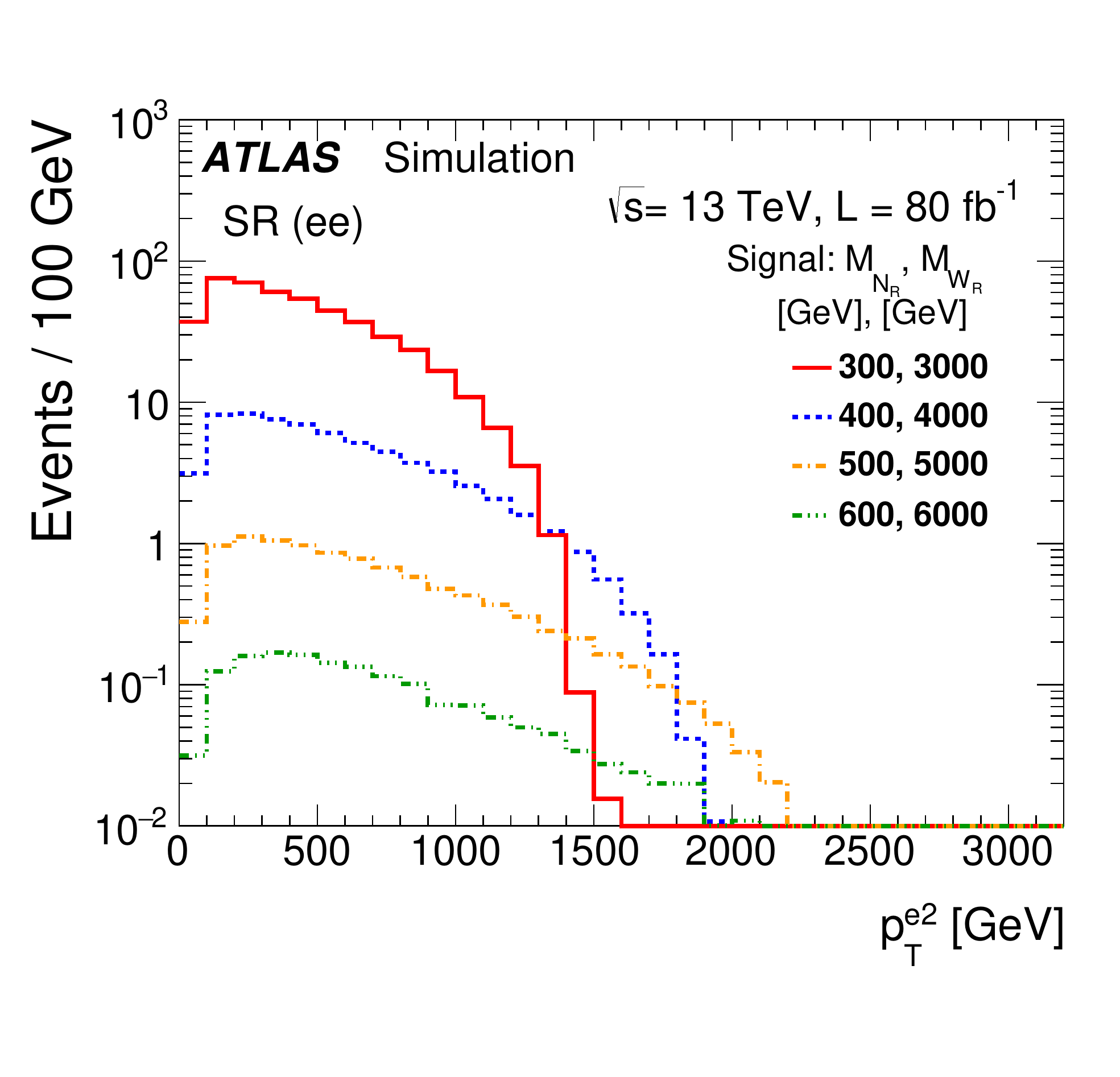} \qquad
\includegraphics[width=0.37\textwidth]{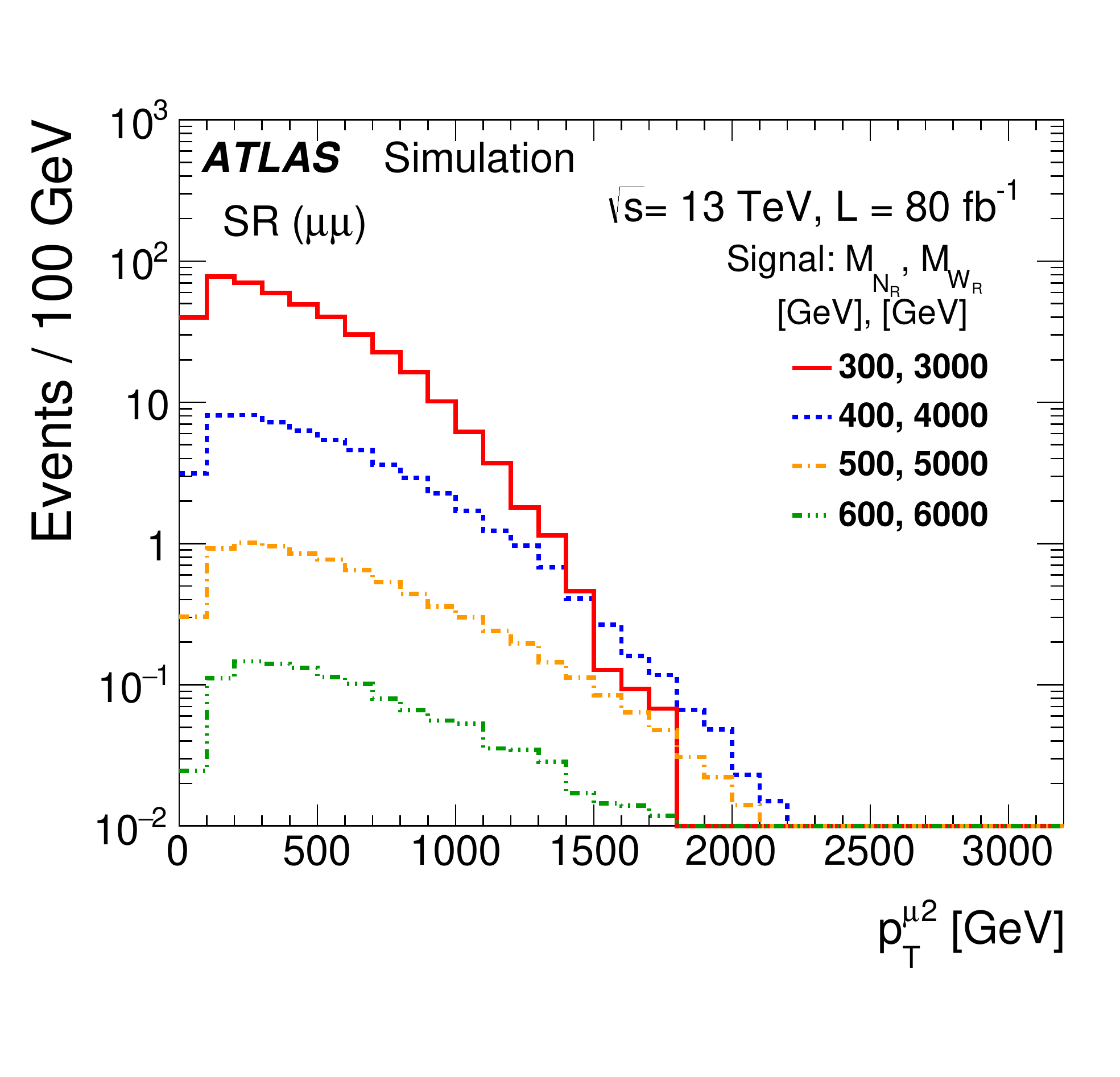} \\[1em]
\vspace{-1cm}
\includegraphics[width=0.37\textwidth]{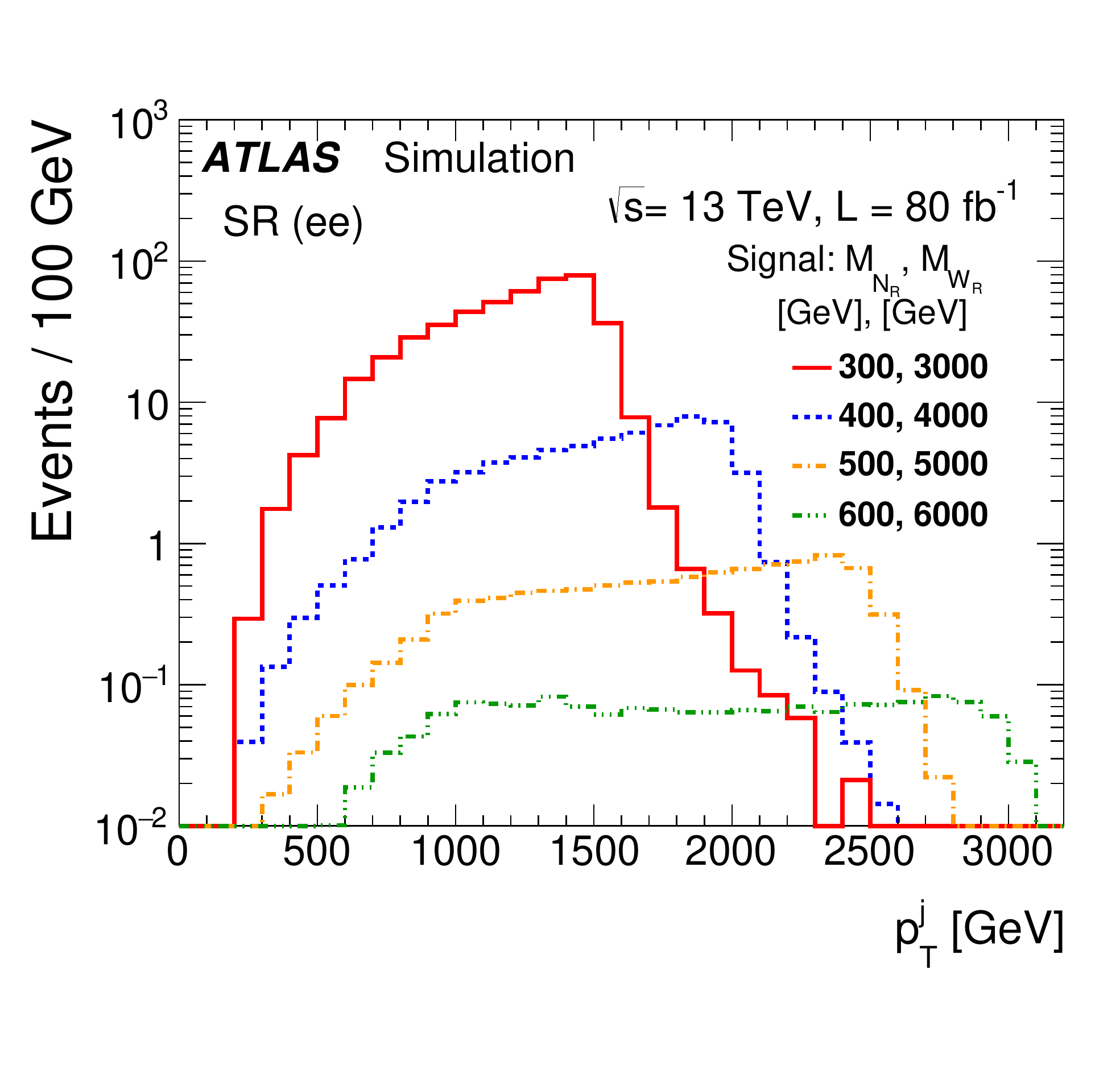} \qquad
\includegraphics[width=0.37\textwidth]{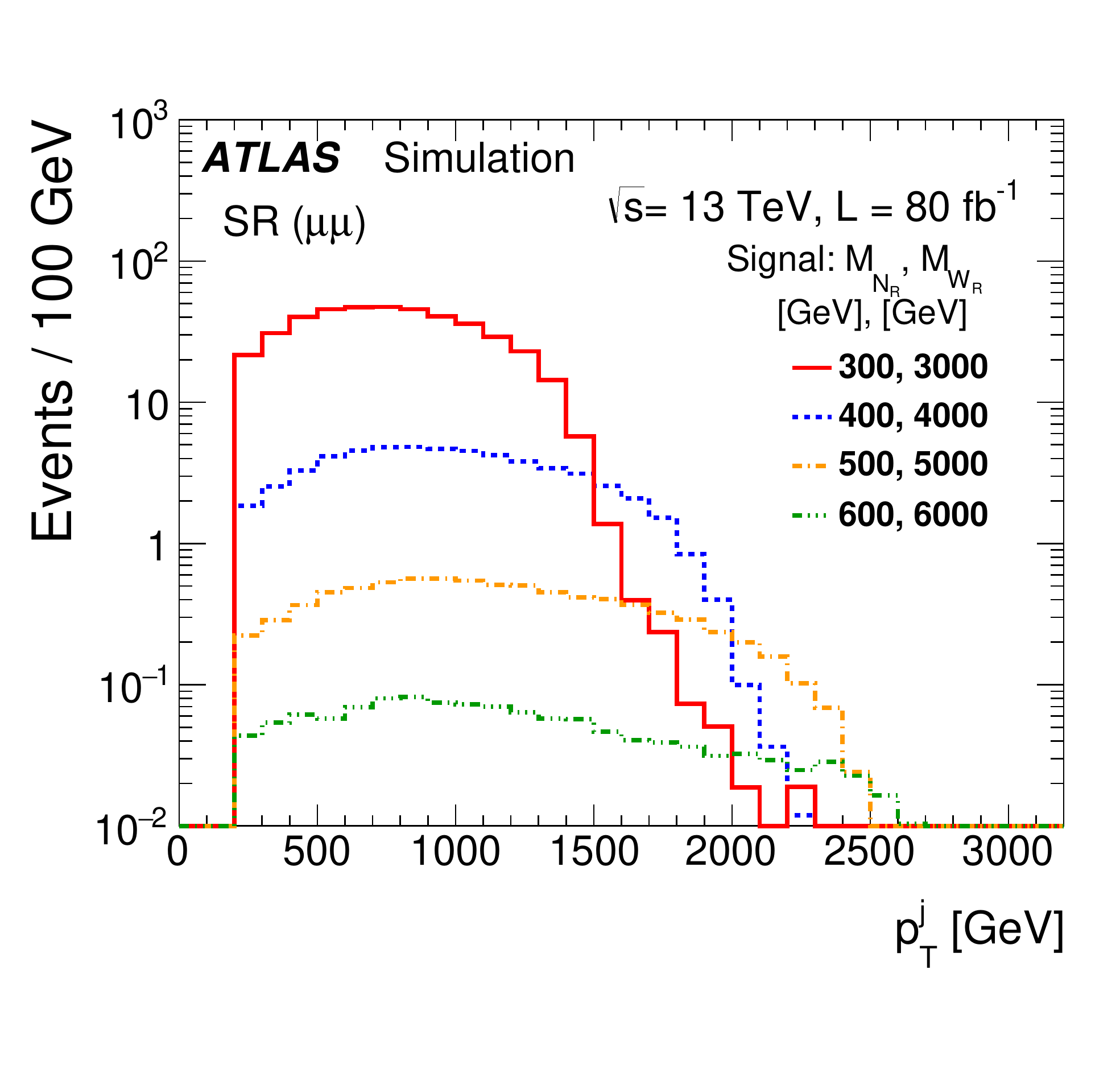} \\[1em]
\vspace{-1cm}
\includegraphics[width=0.37\textwidth]{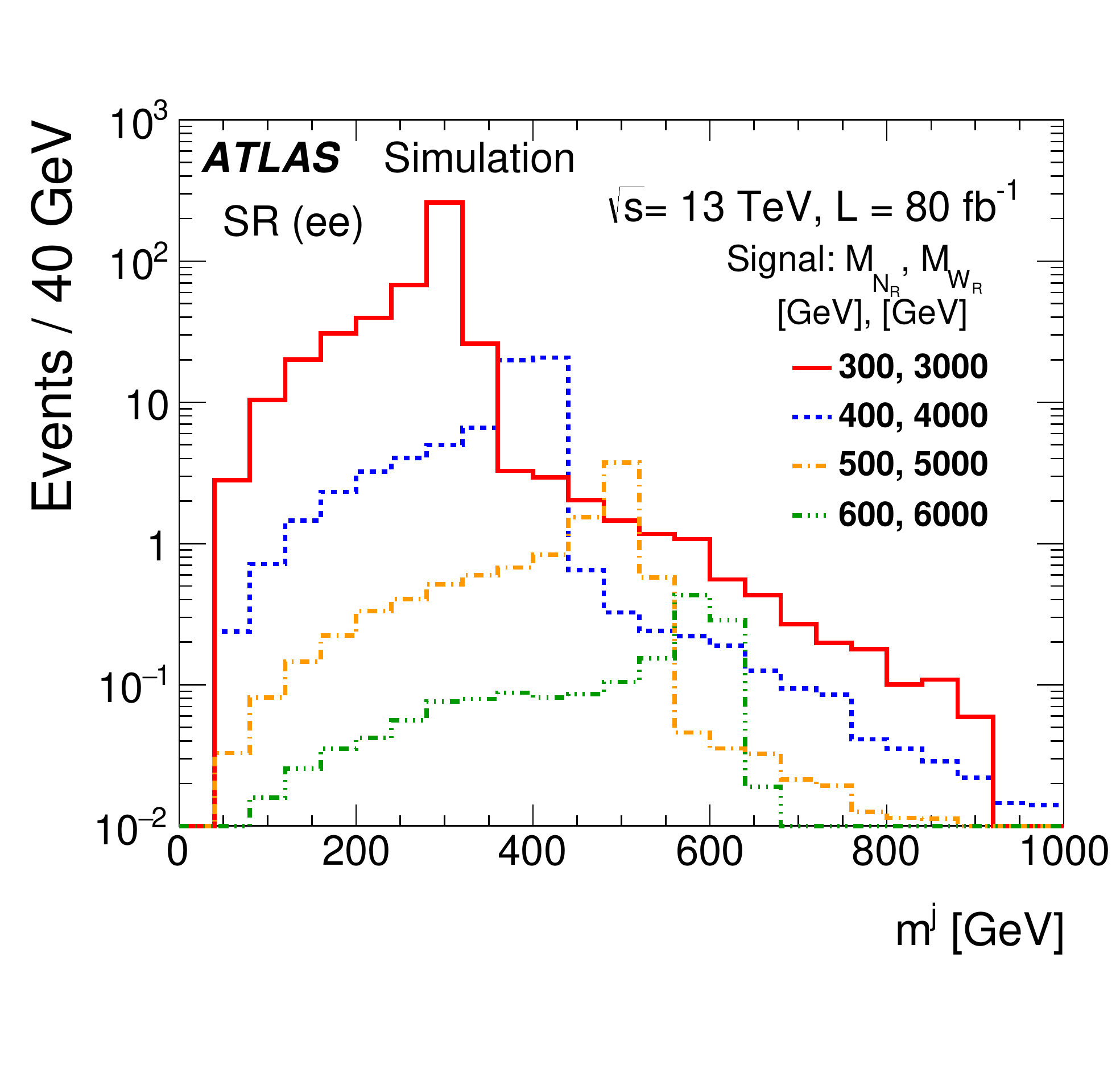} \qquad
\includegraphics[width=0.37\textwidth]{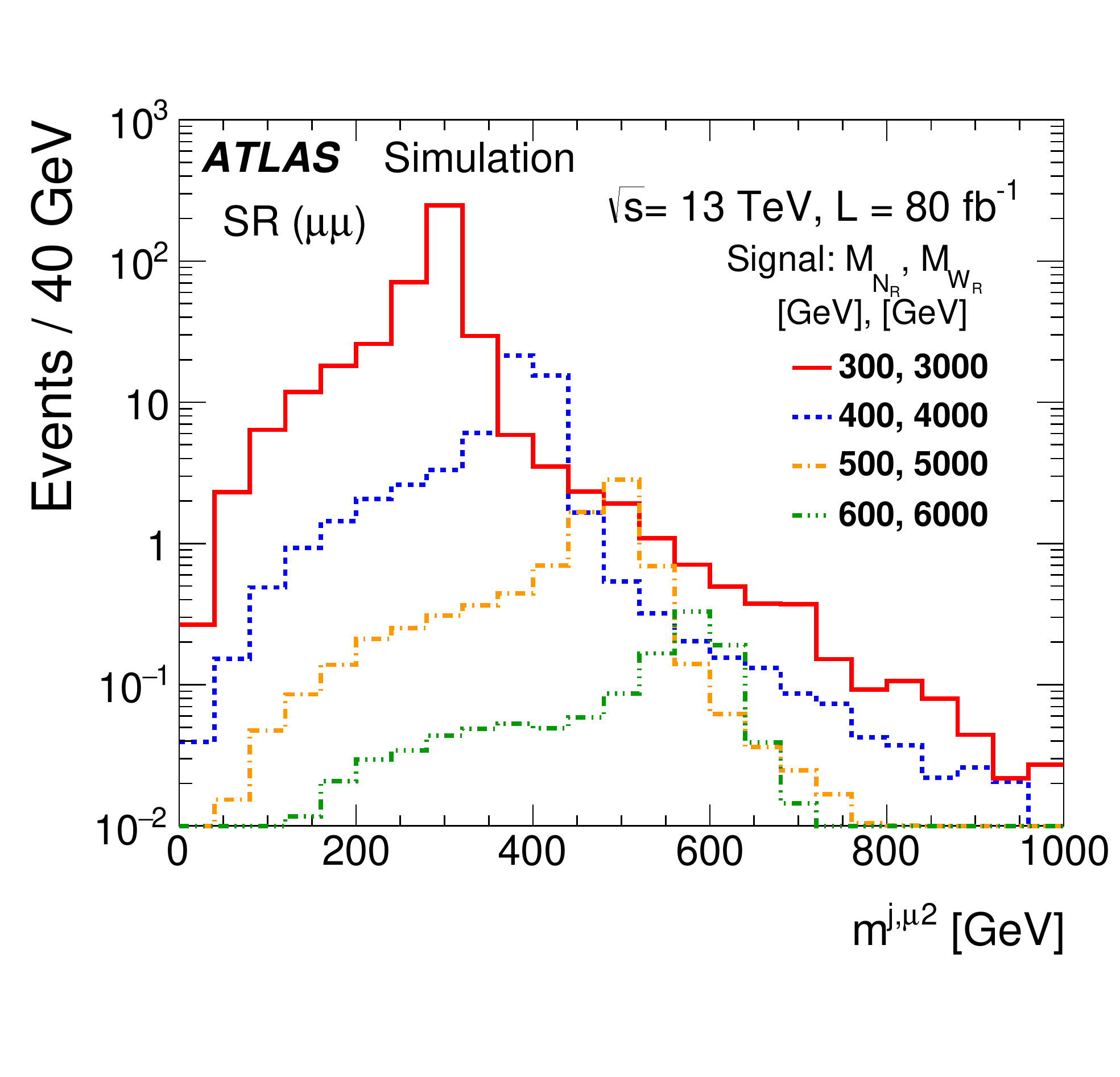} \\[1em]
\vspace{-1cm}
\caption{Reconstructed distributions of the transverse momentum of the leading lepton, subleading lepton,
the selected \LRJ, and the \NR candidate mass in electron (left column) and muon (right column) channels
for four representative signal samples in the signal region.
The indices 1 and 2 indicate leading and subleading lepton, respectively.}
\label{fig:srsignal}
\end{figure}

The event selection is designed to select signal events, while rejecting background events, based on the signal topology.
The events are selected if they contain exactly two same-flavour leptons
(with no charge requirement) and at least one trimmed \LRJ~\cite{Krohn:2009th}
with large transverse momentum \pT$>200$~\GeV.
The highest-\pT (leading) lepton should be back-to-back in azimuth with the \LRJ,
while other (subleading) lepton should be contained in the \LRJ.
In Figure~\ref{fig:srsignal}, the reconstructed \pT distributions of the leading and subleading lepton, as well
as of the selected \LRJ, and the candidate \NR mass are shown for four representative signal samples.
The leading electron and the \LRJ are balanced in \pT, with the maxima at
roughly half of the corresponding \MWR values. The leading muon
\pT shows the same characteristic, but the \pT of the \LRJ is lower and has a broader distribution,
as it does not contain the energy from the subleading muon,
and the muon \pT resolution for high-\pT muons is worse.
The reconstructed mass of the \NR in each case is consistent with the expected value.
The natural width of the resonance varies with the mass and is 100 \GeV\ for $\MWR=3$~\TeV.
At this mass the width of the reconstructed mass peak
is about 150 \GeV\ in the electron channel, and about 350 \GeV\ in the muon channel.

The detailed selection criteria are listed in Table~\ref{tab:obj} and further
discussed below. Events with electrons and muons are analysed
separately.
The leading lepton is required to pass a single-lepton trigger.
For data collected in 2015, the lowest \pT trigger threshold is
24~\GeV\ and 20~\GeV\ for single-electron and single-muon triggers, respectively.
For 2016 and 2017 data, the threshold is 26~\GeV\ for both.

Electron candidates are reconstructed from an isolated energy deposit in the electromagnetic calorimeter
matched to an ID track, within the fiducial region of transverse energy $\pT > 26$~\GeV\ and $|\eta| < 2.47$.
Candidates within the transition region between the barrel and endcap electromagnetic calorimeters,
$1.37 < |\eta| < 1.52$, are excluded.
Muon candidates are reconstructed by combining tracks found in the ID with tracks found in the muon
spectrometer and are required to satisfy $\pT > 28$~\GeV\ and $|\eta| < 2.5$.
Electrons and muons are required to be isolated using criteria based on tracks and calorimeter energy deposits.
For track-based isolation, the discriminating variable is the scalar
sum of the \pT of tracks coming from the primary vertex~\footnote{
Collision vertices are formed from tracks with $pT > 400$~\MeV.
If an event contains more than one vertex candidate, the one
with the highest $\sum \pT^2$ of its associated tracks is selected as the primary vertex.
}
in a variable-size cone around the lepton direction
(excluding the track identified as the lepton), with the
cone size given by the maximum of $\Delta R = 10$~\GeV/$\pT^{\ell}$ and $R_0$, where
$\pT^{\ell}$ is the \pT of the lepton, and $R_0$ is a constant, set to 0.2 for electrons, and 0.3 for muons.
For calorimeter-based isolation, the discriminating variable is the
sum of the transverse energies of topological clusters~\cite{Aad:2016upy} around the lepton in a cone of size $\Delta R  = 0.2$.

The inputs to the jet construction are noise-suppressed three-dimensional topological clusters of energy deposits in the calorimeters,
built from calorimeter cells~\cite{Aad:2016upy}.
They are classified as either electromagnetic or hadronic, based on their shape, depth and energy density.
The energy clusters are calibrated to the hadronic scale.
The momenta of the jets are corrected for energy losses in passive material and for the non-compensating
response of the calorimeter~\cite{Aaboud:2018kfi}.
The \LRJS are constructed with the \antikt algorithm~\cite{Cacciari:2008gp} with a radius parameter of $R=1.0$,
through its implementation in FastJet~\cite{Cacciari:2011ma}.
They are further trimmed~\cite{Krohn:2009th} to reduce the contamination from
soft uncorrelated radiation.
In this method,  the original constituents of the jets are reclustered using
the \kt
algorithm~\cite{Catani:1993hr} with a radius parameter $R_{\textrm{sub}} = 0.2$
in order to produce a collection of subjets.
These subjets are discarded if they carry
less then a specific fraction ($f_{\textrm{cut}} = 5\%$) of the \pT of the original jet.
The remaining constituents are summed to form the four-momentum of the final jet.
 
In the electron channel, the \LRJS are required to have a mass of at least $50$~\GeV, while
no such requirement is applied in the muon channel. This is because in the former case,
the \LRJ includes the electron,
while in the muon channel, the muon is not included in the \LRJ.
 
\begin{table}[tbp]
\caption{Object selection criteria.
The significance of the transverse impact parameter is
defined as the transverse impact parameter $d_0$ divided by its
uncertainty, $\sigma_{d_0}$, of tracks relative
to the primary vertex with the highest sum of track \pT.
The longitudinal impact parameter $z_0$
is multiplied by $\sin\theta$, where $\theta$ is the polar angle of  the track.}
\centerline{
\begin{tabular}{lll}
\toprule
&  Electron channel  &  Muon channel \\
\midrule
\multicolumn{3}{l}{\textbf{Lepton:}} \\
\midrule
\pT    & > 26 \GeV  & > 28 \GeV \\
$|\eta|$ & $|\eta| < 1.37$ or $1.52<|\eta|<2.47$ & $< 2.5$ \\
Leading lepton quality & Medium~\cite{Aaboud:2019ynx}, isolated~\cite{Aaboud:2019ynx} & Medium~\cite{Aad:2016jkr}, isolated~\cite{Aad:2016jkr} \\
Subleading lepton quality &  Medium, no isolation & Medium, no isolation\\
Transverse impact parameter significance & $|d_0|/\sigma_{d_0} < 5.0$ &  $|d_0|/\sigma_{d_0} < 3.0$\\
Longitudinal impact parameter & \multicolumn{2}{c}{$|z_0|\sin\theta <$ 0.5 mm} \\
\midrule
\multicolumn{3}{l}{\textbf{Trimmed \LRJ:}} \\
\midrule
\pT  & \multicolumn{2}{c}{> 200 \GeV} \\
$|\eta|$ &\multicolumn{2}{c}{< 2.0 }\\
Mass & $>50$~GeV & None \\
\bottomrule
\end{tabular}}
\label{tab:obj}
\end{table}

Small-radius jets constructed with the
\akt algorithm using energy clusters calibrated to the electromagnetic scale
with a radius parameter of $R = 0.4$ are used
to check for possible overlap between
objects, and to perform \btagging (described in Section~\ref{sec:elinjet}).
In the muon channel, the event is discarded if either muon
satisfies $\Delta R_{\mathrm{y}}  (\mu,\textrm{jet}) < min(0.4, 0.04 + 10$~\GeV/$p_{\textrm{T}}^{\mu})$,
in order to avoid jets formed from energy deposits associated to high energy muons.
In the electron channel, for the leading electron, first all small-radius jets
within $\Delta R_{\mathrm{y}}  = 0.2$ of a
selected electron are removed.
Then the event is discarded if
the leading electron is within $\Delta R_{\mathrm{y}}  = 0.4$
of a remaining small-radius jet.
This is referred to as the nominal overlap
removal procedure for electrons.
A modified procedure, described in Section~\ref{sec:elinjet},
is applied for the subleading electron as, unlike muons,
electron clusters can overlap with a jet and the signal efficiency drops off if the standard overlap removal approach is followed.

Further requirements based on the characteristics of the signal are applied:
 
\begin{itemize}
 
\item $\Delta R_{\mathrm{y}} $ between the subleading lepton and the \LRJ is required be less than 1, in order for
the lepton to be inside the jet.
 
\item $\Delta \phi$ between the leading lepton and the \LRJ is required to be greater than 2, in order to
have a balanced topology between the \NR and the lepton from the \WR decay.
 
\item In order to reduce the $Z$+jets background, events with a dilepton invariant mass of less than 200~\GeV\ are
vetoed, and the $\Delta \phi$ between the leading and subleading leptons is required to be greater than $1.5$.
 
\end{itemize}
 
After applying these requirements, simulation studies show that the background consists mainly of $t\overline{t}$
and $Z$+jets processes (including off-shell $Z/\gamma^{*}$ production),
while  contributions from $W$+jets, single-top-quark and multijet processes are negligible.
No requirements on \btagged jets are applied, as the \WR in the signal can decay to a top and bottom quark pair.

The final discriminating observable used in the analysis is the reconstructed mass of the \WR candidate, \MWRR.
In the electron channel, the selected \LRJ corresponds to the \NR candidate, and therefore the \WR candidate four-momentum is obtained
by adding the \LRJ and the leading electron four-momenta. In the muon channel, the \NR candidate
four-momentum is obtained by adding the four-momentum of the selected \LRJ to that of the muon contained in the jet.
The \WR candidate four-momentum is obtained by adding the \NR candidate four-momentum to that of the leading muon.
In both cases, if there is more than one \LRJ in the event, the \LRJ with the largest mass is used.
 
Based on the range of \MWRR, control and signal regions (CR, SR) are defined as specified in Table~\ref{tab:reg}.
The CR is defined in a region of low reconstructed \MWRR corresponding to a parameter space
excluded by previous searches~\cite{Aaboud:2018spl}.
The background in the SR is evaluated from
a combined fit of MC and data events in the CR (described in Section~\ref{sec:bg}).
To test the performance of \LRJS containing electrons, a validation region (VR)
is defined with a selection orthogonal to the CR and the SR.
This requires a muon balanced in \pT by a \LRJ with an electron inside.
By construction, the VR is dominated by
\ttbar events decaying dileptonically
to $e\mu$ final states.

\begin{figure}[h!]
\centering
\includegraphics[width=0.45\textwidth]{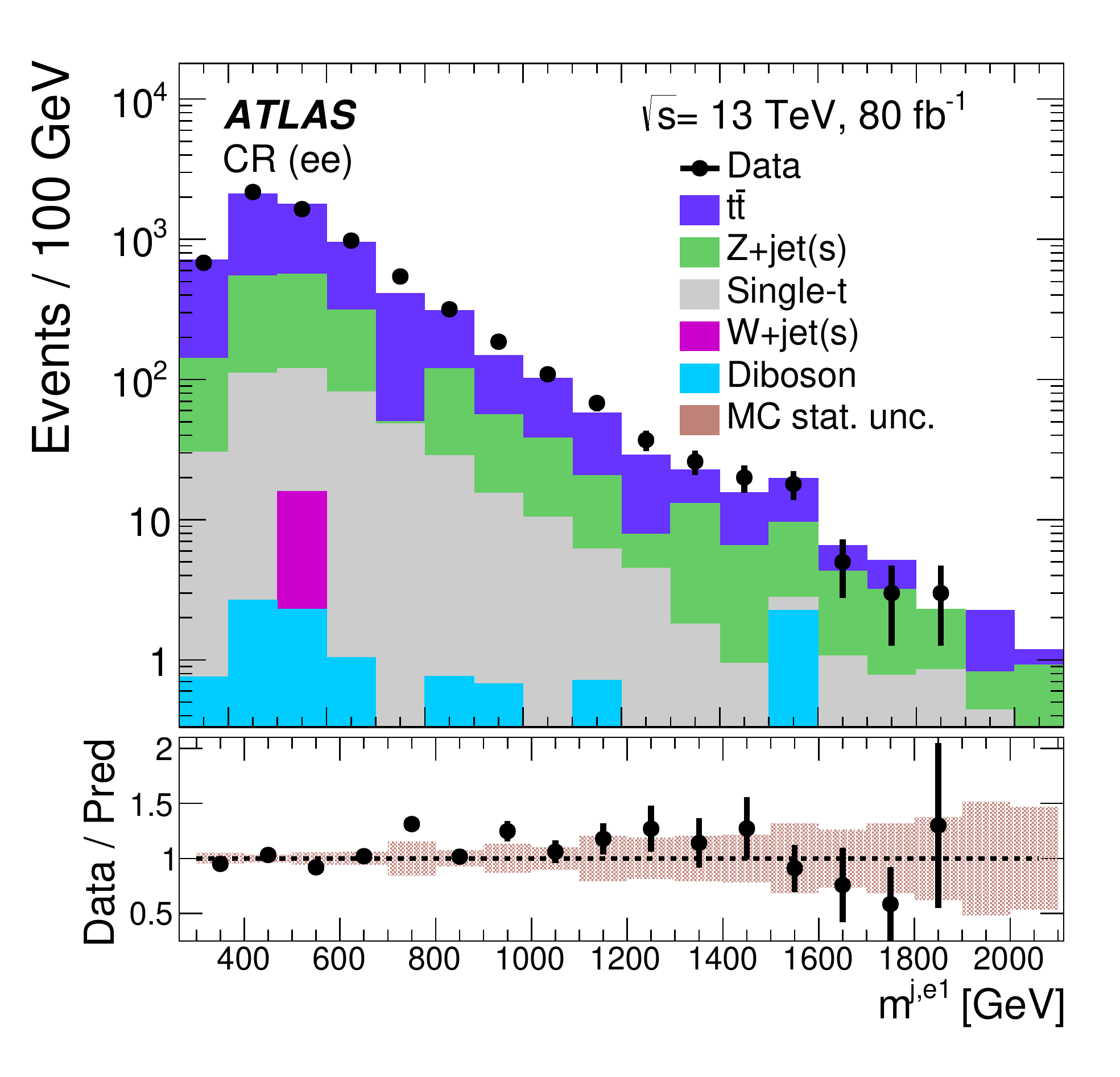} \qquad
\includegraphics[width=0.45\textwidth]{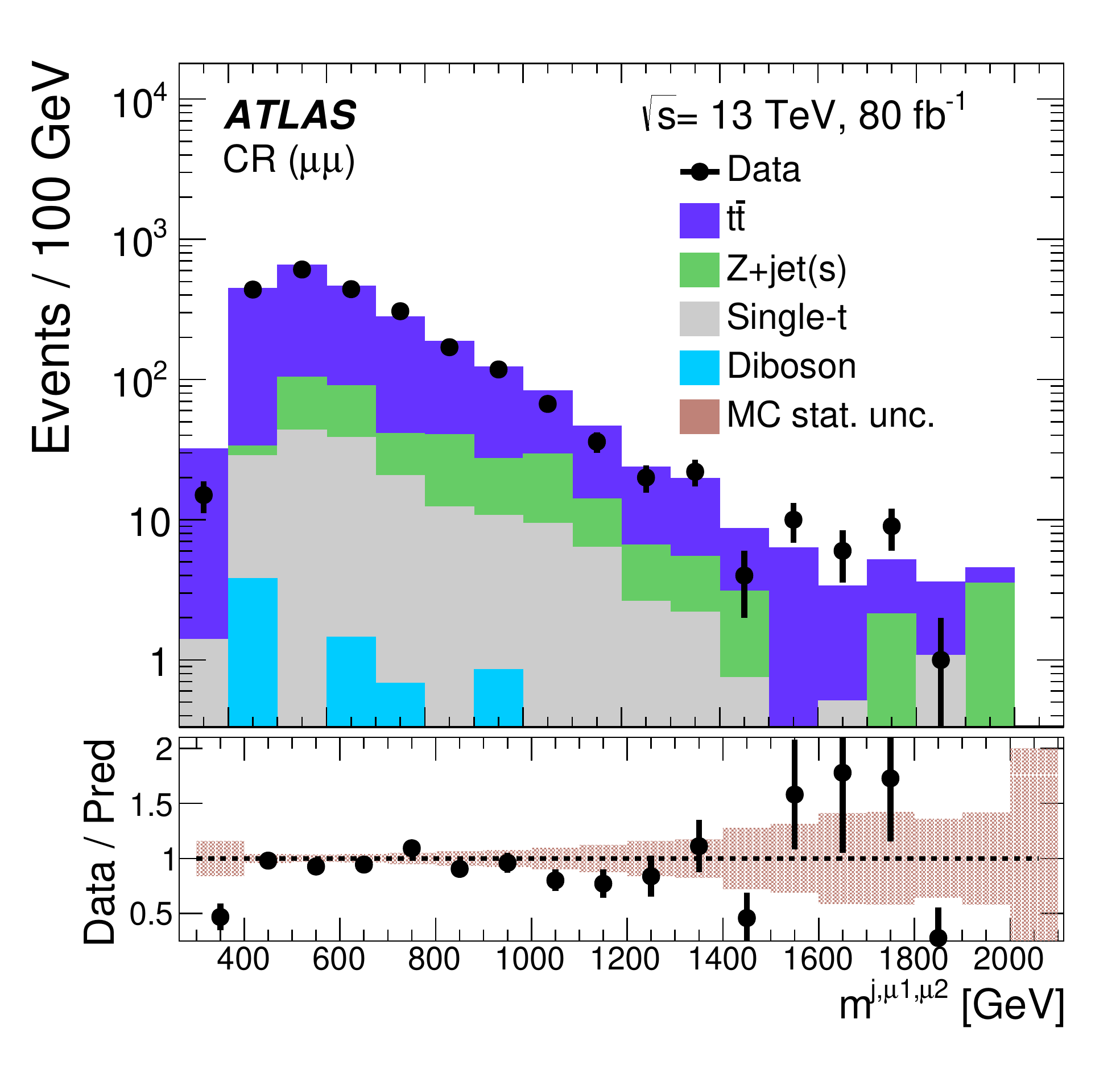} \\[1em]
\includegraphics[width=0.45\textwidth]{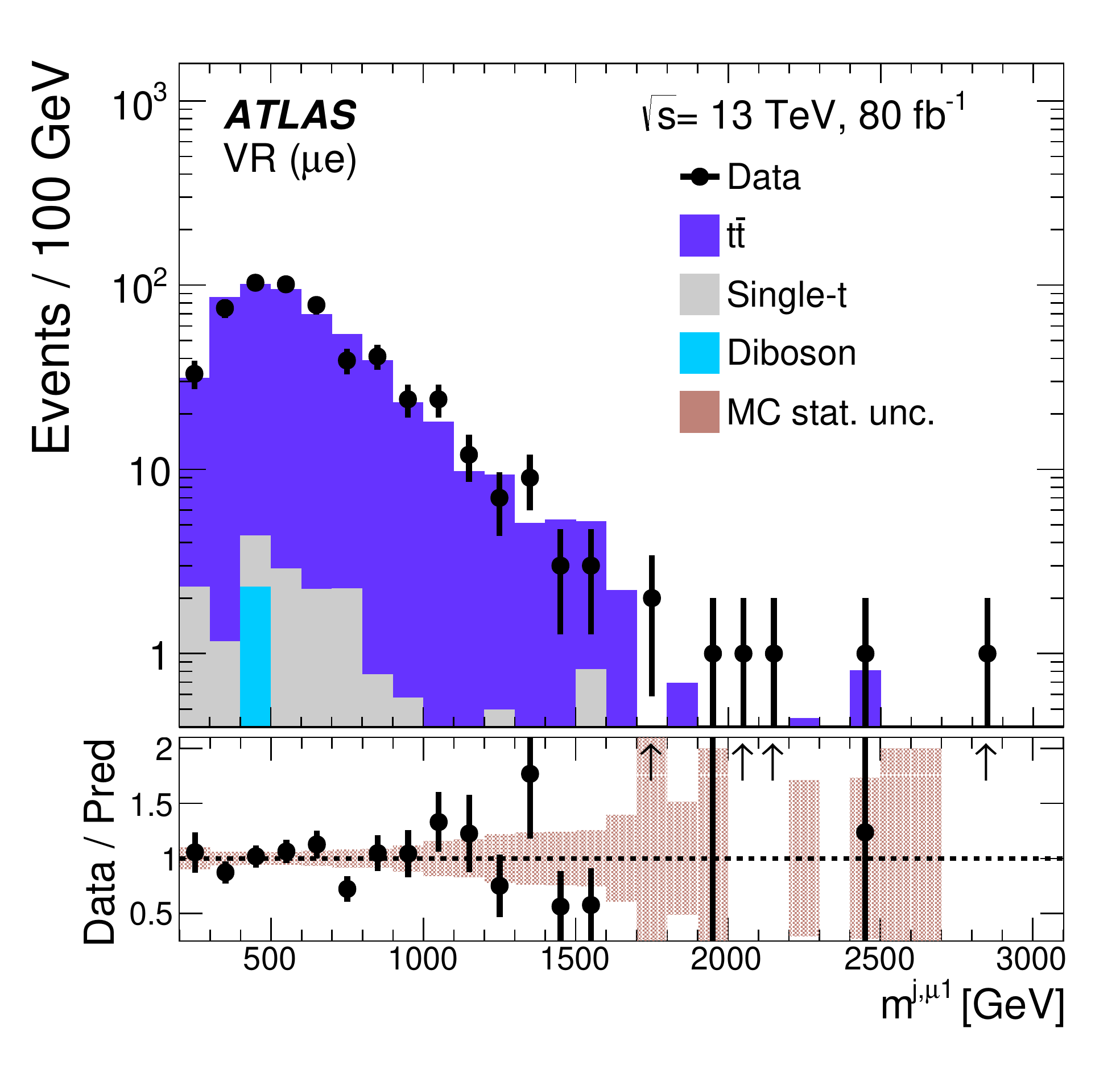} \qquad
\includegraphics[width=0.45\textwidth]{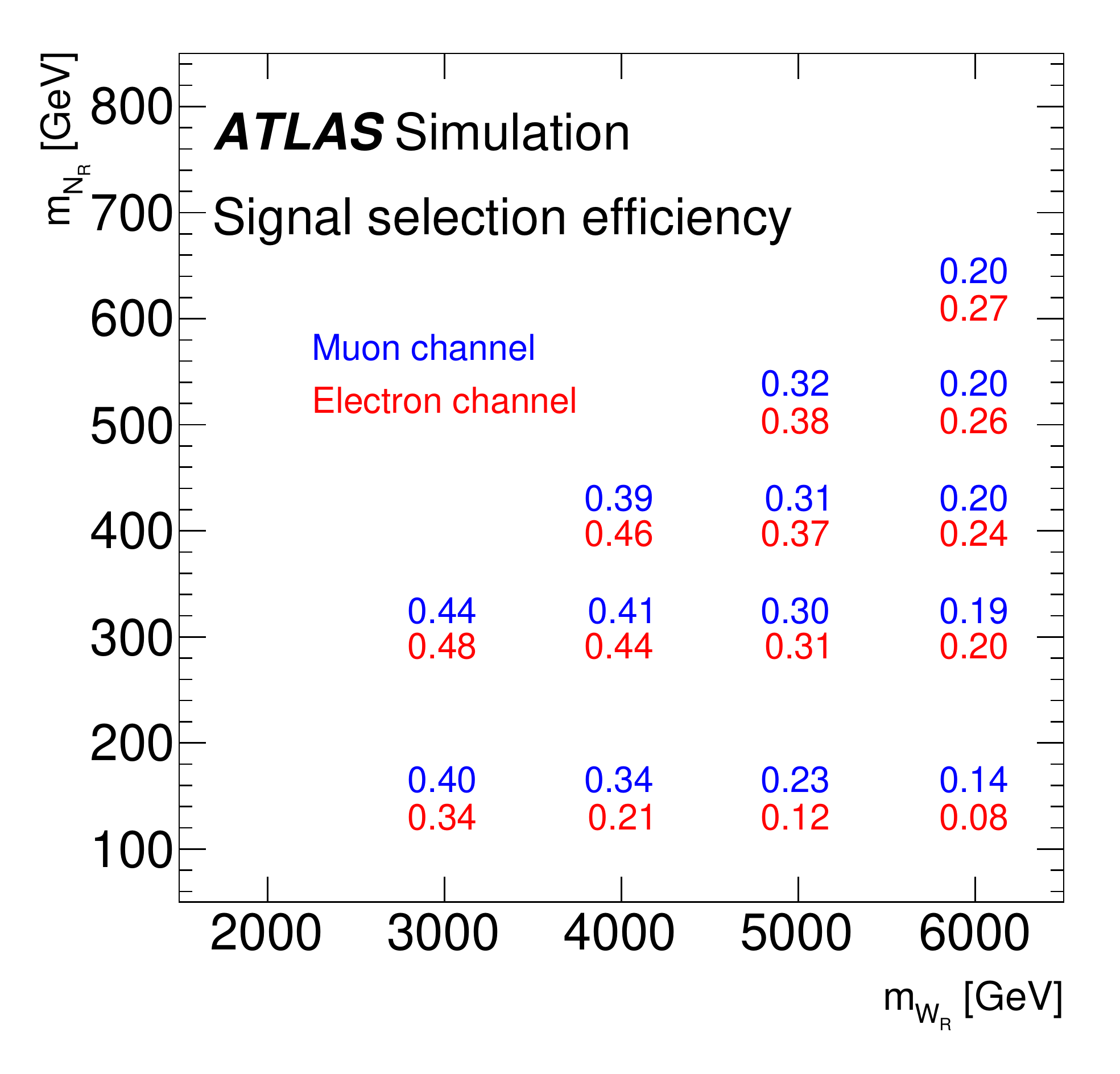} \\[1em]
\caption{Distributions of the reconstructed mass of the \WR, in the electron (top left), muon (top right) channel control
regions, in the validation region (bottom left),
and the signal selection efficiency for a number of benchmark points (bottom right) using simulated signal samples.
The \MWRR distributions only include statistical uncertainties, which are shown for both the data and simulation
in the ratios. }
\label{fig:WRM}
\end{figure}

In Figure~\ref{fig:WRM},
good agreement is observed between data and
simulation in the \MWRR distributions in the control regions of the electron and muon channels,
as well as in the validation region.
In the bottom-right plot, the selection efficiency times acceptance is shown
for different signal samples.
The efficiency decreases for lower \MNR and higher \MWR values.
The largest inefficiency arises from the difficulty of electron reconstruction
close to hadronic activity, which is discussed in the next section.
The probability of producing an off-shell \WR increases with the mass.
This can result in a less boosted \NR, explaining the drop in signal efficiency for higher \MWR values.

\begin{table}[tbp]
\caption{Definition of signal, control and validation regions}
\centerline{
\begin{tabular}{lll}
\toprule
Region & Range of \MWRR &  Lepton flavour  \\
\midrule
Signal region (SR)  & $ > 2$ TeV & Same flavour \\
Control region (CR)  &  $ < 2$ TeV & Same  flavour \\
Validation region (VR)  & All & Mixed flavour (leading: muon; subleading: electron) \\
\bottomrule
\end{tabular}}
\label{tab:reg}
\end{table}
 
\FloatBarrier
 
\section{Performance of \LRJS containing electrons}
\label{sec:elinjet}
 
A distinguishing feature of this search is the use of
\LRJS containing electrons as a proxy for \NR in the electron channel.
Since the \LRJ construction
procedure is based on energy clusters calibrated at the hadronic scale, the effect
of a non-negligible fraction of EM clusters in the \LRJ needs to be investigated.
The analysis does not use the kinematic properties of the identified electron inside
the \LRJS to reconstruct the \NR or \WR invariant masses,
but uses the mass of the \LRJ, which includes the associated electron clusters.
The presence of real hadronic activity close to an electron may
affect the reconstruction of the electron.

\begin{figure}[h]
\centering
\includegraphics[width=0.45\textwidth]{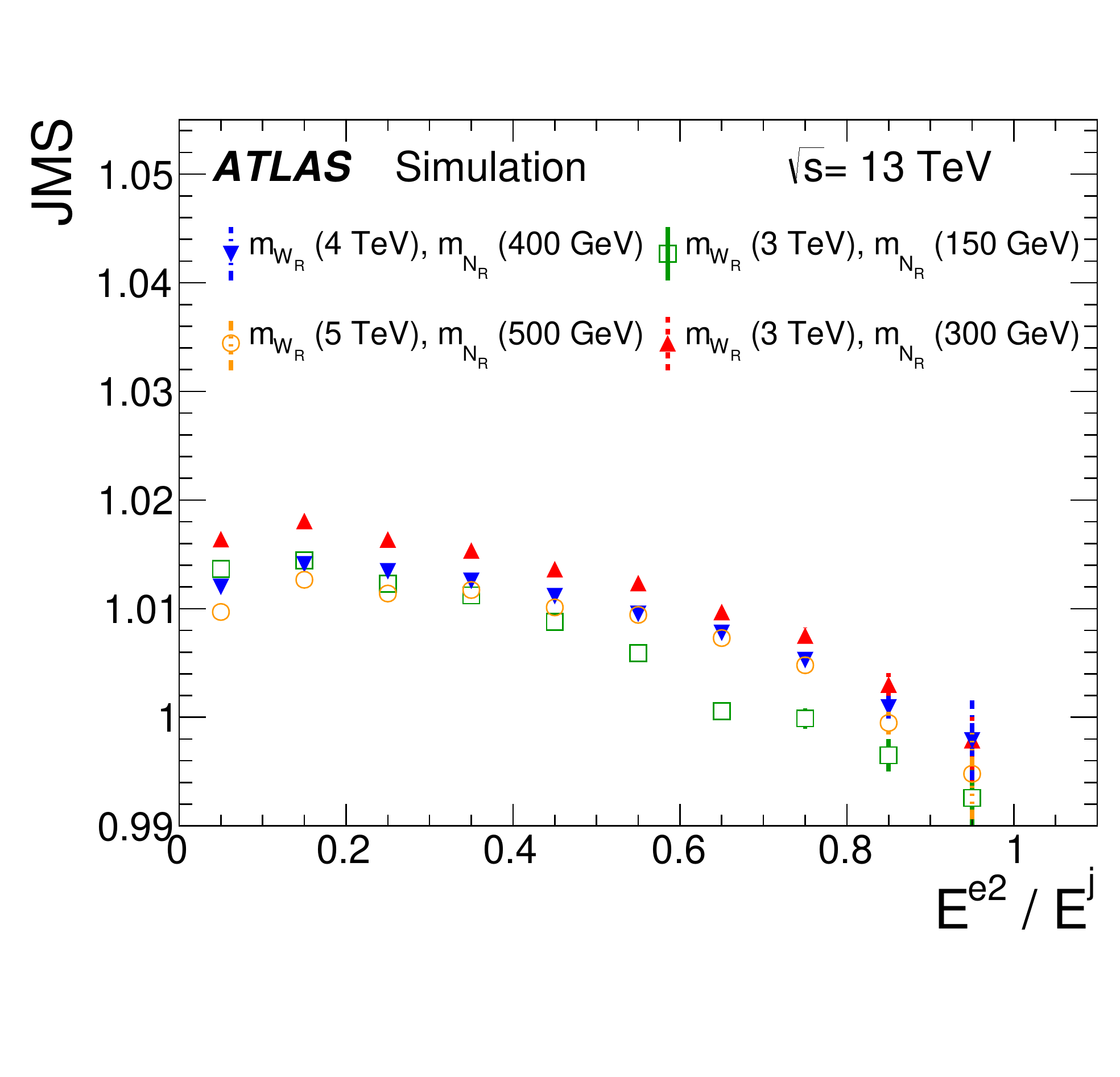}\qquad
\includegraphics[width=0.45\textwidth]{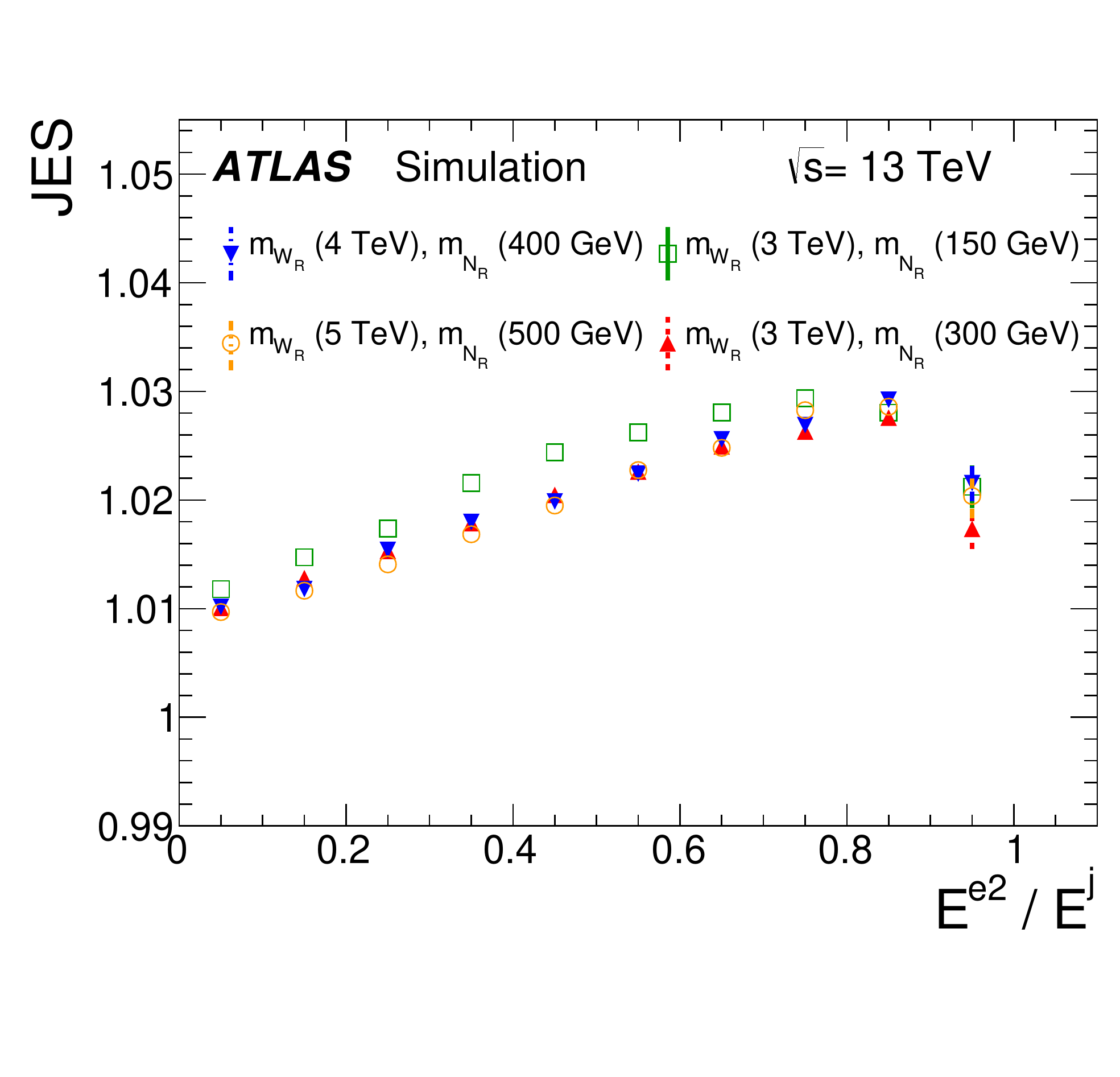}\\[1em]
\caption{Large-$R$ jet average JMS (left) and JES (right) as a function of subleading electron energy divided by the \LRJ energy
for signal samples.
}
\label{fig:elen}
\end{figure}

The jet mass and energy scales, JMS and JES, defined as the average of the ratio
of the mass or energy of the reconstructed and corresponding
generator-level \LRJS, are used to study the
effect of including the large EM-cluster of the electron in the jet
reconstruction.
The matching between detector-level and
generator-level \LRJS is performed with $\Delta R_{\mathrm{y}}  < 0.75$.
The generator-level jet
is obtained by clustering stable final-state particles (with lifetime greater than 30 ps) except muons and neutrinos using
the same jet algorithm, radius parameter and trimming used at the detector-level.
The JMS and JES  of the selected \LRJS for a few representative signal samples are shown in Figure~\ref{fig:elen}
as a function of the ratio of the energy of the electron to the energy of the \LRJ. This ratio can be considered a proxy for the electromagnetic
energy fraction in the \LRJ.
Constant values of JES and JMS
within a few percent of unity indicate that the \LRJ
has only a weak dependence on the fraction of electromagnetic energy inside the jet,
and thus no particular additional corrections are required for the signal \LRJS.
Typical numbers for the \LRJ mass resolution (JMR) in signal events are about 4-6\%
in the electron channel and about 7-14\% in the muon channel,
while the \LRJ energy resolution (JER) is about 3-5\% GeV in both channels.
As opposed to the muon channel, in the electron channel the \LRJ does contain the electron as a compact and high energy deposit.
This implies a more precise angular distribution of the energies in the jet and thus a better JMR in the electron channel.

In signal events, as almost all selected \LRJS contain activity from
both the \WR hadronic decay and the nearby electron,
application of the nominal overlap removal procedure, as described in Section~\ref{sec:sel}, for the subleading electron
results in a large loss of signal efficiency.
However, it is observed that removing events where the electron and the nearest jet are within
$\Delta R_{\mathrm{y}}  < 0.04$ retains a sizeable
fraction of the signal events and rejects a large fraction of the background events.
The discarded events are dominated by the case where isolated electrons are reconstructed as jets.

A study is performed to check the validity of the
electron efficiency correction factors, which account for potential differences in electron
reconstruction, identification and isolation
efficiency between data and simulation and are derived using well-isolated electrons~\cite{Aaboud:2019ynx}.
A sample of $t\overline{t}$ events decaying into a mixed-flavour dileptonic final state is chosen.
The leading lepton is chosen to be a muon with nominal
isolation requirements, and the subleading lepton is chosen to be an electron with no isolation requirements.
The rest of the selection is the same as the signal selection, except that the events are selected with at least one $b$-tagged
small-radius jet.
The \btagging is based on a multivariate algorithm~\cite{Aad:2015ydr}.
Several observables based on the long lifetime
of $b$-hadrons and the $b$- to $c$-hadron decay topology, are used as algorithm input to discriminate between
$b$-jets, $c$-jets and other jets. The \btagging requirement corresponds to the
$70\%$ efficient working point for $b$-jets, as determined in simulated \ttbar\ events, while the
rejection rates of $\tau$-jets, $c$-jets and light-flavour jets are
$55$, $12$ and $381$, respectively~\cite{ATL-PHYS-PUB-2015-022, ATL-PHYS-PUB-2016-012}.

\begin{figure}[h]
\centering
\includegraphics[width=0.45\textwidth]{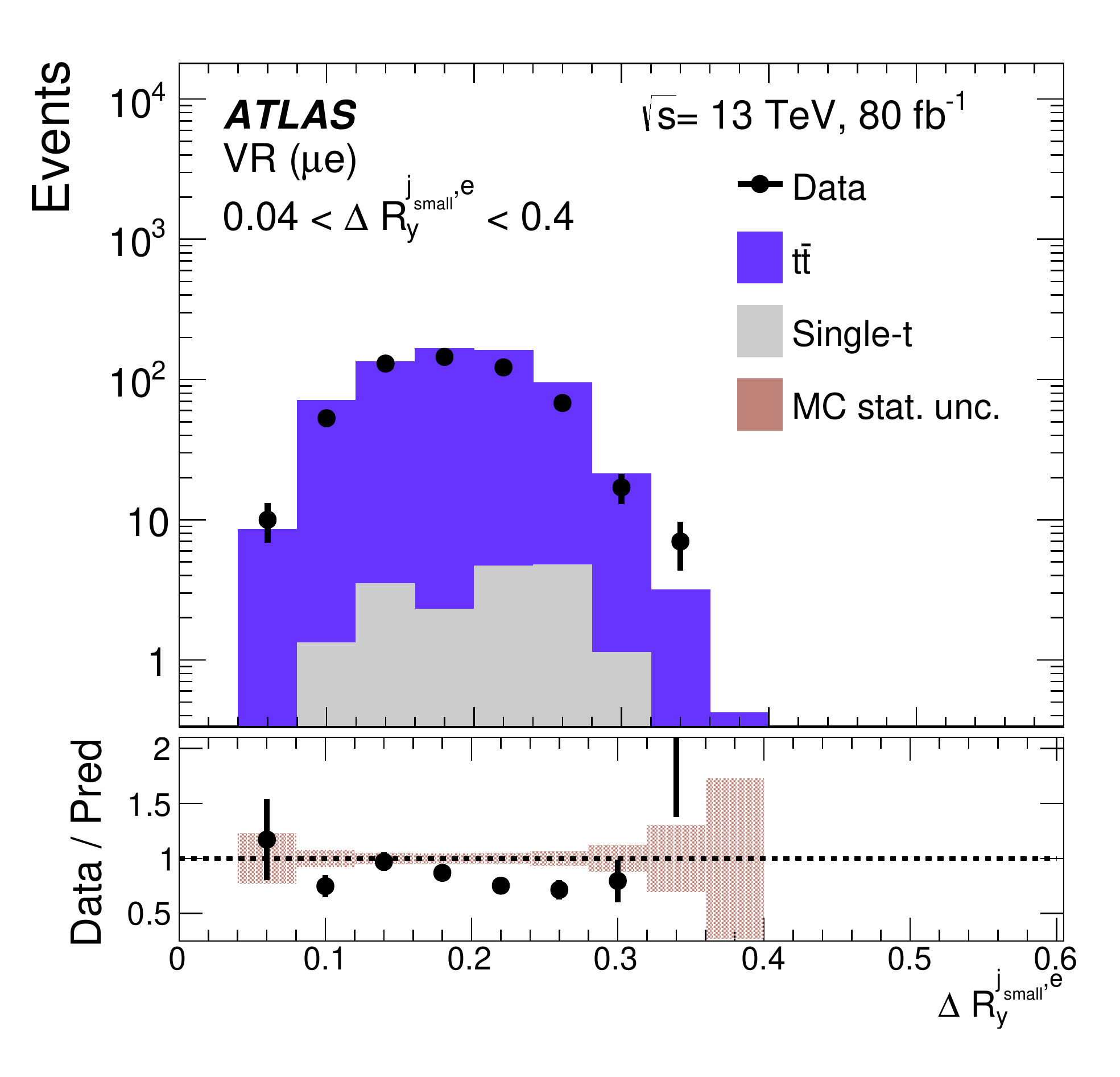}\qquad
\includegraphics[width=0.45\textwidth]{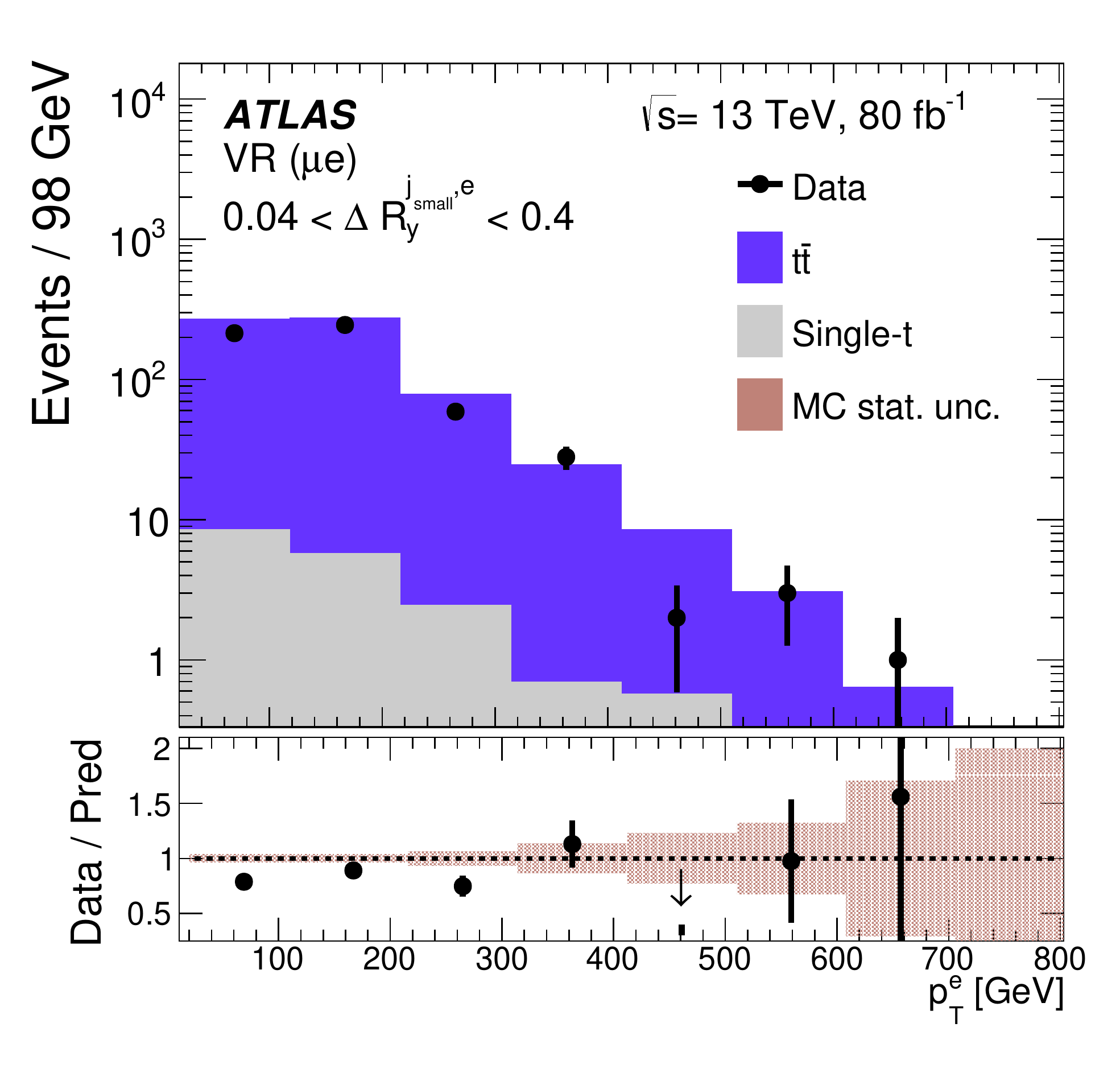}\\[1em]
\caption{
Angular separation between the subleading electron and nearest small-radius jet (left) and
transverse momentum distribution of the subleading electron (right)
in events with a leading muon and subleading electron, requiring one \btagged small-radius jet.
The distributions only include statistical uncertainties, which are shown for both the data and simulation
in the ratios.
}
\label{fig:eljetkin}
\end{figure}

The \LRJ requirements are the same as in the nominal selection.
Only the events in which the electrons and the nearest small-radius jet are within
$0.04 < \Delta R_{\mathrm{y}}  < 0.4$ are studied.
The distributions of electron \pT and $\Delta R_{\mathrm{y}}$ between the electron
and the small-radius jet show satisfactory agreement between data and simulation, as shown in Figure~\ref{fig:eljetkin}.
An additional uncertainty of 30\% on the electron
efficiency for electrons within $0.04 < \Delta R_{\mathrm{y}}  < 0.4$ is
derived from the difference in yields between data and the simulation,
statistical uncertainties as well as the theoretical and \btagging
uncertainties on the simulation.

\FloatBarrier

\section{Background estimation and systematic uncertainty}
\label{sec:bg}
 
While the CR is dominated by \ttbar events,
the fraction of $Z$+jets events becomes larger at higher masses as they have a less steeply falling
mass distribution than \ttbar events, as shown
in Figure~\ref{fig:WRM}.
In order to take this into account in evaluating the background in the SR,
which is located at higher masses, a data-driven approach is used to evaluate the \ttbar contribution in
conjunction with a fitted MC prediction of the $Z$+jets background. The fit to the data in the CR is extrapolated to the SR.
 
Different steeply falling functions are tested,
motivated by Refs.~\cite{Aad:2015mzg, Aaboud:2017yvp},
where they are found to
fit steeply falling distributions like the scalar sum of
jet transverse momenta in multijet events or the dijet mass.
It is observed that the functional form $A \exp{(-Bu)}/u^C$
describes the data distribution in the CR the best, while the functional form
$A^{\prime} (1-u)^{B^{\prime}} (1+u)^{(C^{\prime}u)}$ best describes the
$Z$+jets MC distribution. In both cases, $u = \MWRR/\sqrt{s}$
and
the choice of fit function is determined by the goodness of fit (based on the $\chi^2$ per
degree of freedom) as well
as by the best agreement with the yields from the MC background estimate in the CR.
First, the fit parameters $A^{\prime}$, $B^{\prime}$, and $C^{\prime}$
are determined from $Z$+jets MC using a reconstructed \MWRR range of 400--4000~\GeV, then
the fit to data is performed using the function
$A \exp{(-Bu)}/u^C +A^{\prime} (1-u)^{B^{\prime}} (1+u)^{(C^{\prime}u)}$, to determine
the values of the parameters $A$, $B$ and $C$.
This functional form is fitted in the CR range of 600--1800~\GeV, where the range is chosen
depending on the goodness of the fits.
The slope of the background fit is steeper in the muon channel compared to the electron channel.

In the VR the fit performed for reconstructed \WR candidate masses
600--1800 \GeV\ is extrapolated into the region $>2$~\TeV\ and the fit
prediction is compared to the data yield finding consistency.
In order to assess the systematic uncertainty related to the \ttbar data-driven fit,
variations of the data fit range are made,
and the largest change in the SR yield obtained from these variations is taken as the uncertainty.
The same is done for the $Z$+jets MC fit,
and the uncertainty is added in quadrature to the uncertainty derived from fitting alternative $Z$+jets MC samples
obtained after varying the scale (by factors of two and one half) and using alternative PDF sets~\cite{Aaboud:2018spl}.
This uncertainty is larger than the difference yield obtained using \powheg+\pythiaeight.
The relative uncertainty of the background yield in the SR is about $25\%$ for both channels.
Statistical uncertainties on the fits are
estimated using pseudo-experiments built by varying the input data points
within their statistical uncertainties.
The resultant background fit, along with
the estimated uncertainty is shown in Figure~\ref{fig:fit-final}.

\section{Systematic uncertainties of the signal yield}
\label{sec:syst}
 
Systematic uncertainties affect the shape and normalisation of the \MWRR distributions,
thereby changing the signal yield.
The dominant uncertainties in the SR are shown in Table~\ref{tab:syst}.
They can be classified as originating from experimental or theoretical sources.
 
The yields from simulated samples are affected by uncertainties related to the description of the detector response.
The dominant uncertainty is related to
the electron and muon identification, which is (4--20)\%
in the electron channel (including the additional
$30\%$ uncertainty for electrons reconstructed
nearby a small-radius jet) and (4--8)\% in the muon channel, depending on the values
of \MWR and \MNR. The uncertainties related to the electron trigger, reconstruction, and isolation are (4--5)\%.
The uncertainties related to the muon trigger and isolation are about $1\%$.
These uncertainties are assessed by comparing data and simulation
samples of $Z \to \ell^{+} \ell^{-}$ decays~\cite{Aad:2016jkr, Aaboud:2019ynx}.
The simulation is used to extrapolate to lepton \pT beyond a few hundred \GeV.
The other uncertainties such as those related to JES and JMS of the \LRJS
are evaluated by comparing the ratio of calorimeter-based to track-based measurements in multijet events in data and
simulation~\cite{Aad:2013gja, Aaboud:2017jcu, Aaboud:2018kfi}.
The uncertainties related to JMR and JER modelling are evaluated by increasing the nominal resolution by 20\%~\cite{ATL-PHYS-PUB-2018-015}
and 2\% respectively.
These uncertainties are at sub-percent level.
The average number of interactions per bunch crossing is
rescaled to improve the agreement of simulation with data, and the corresponding uncertainty,
as large as the correction, has an effect of $0.5\%$ in the electron channel, and up to $3\%$ in the muon channel,
caused by the lack of a \LRJ mass threshold in the latter case.
The uncertainty in the 2015, 2016 and 2017 integrated luminosity is 2\%.
It is derived from the calibration of the luminosity scale using $x$--$y$ beam-separation scans, following a methodology similar to that detailed in Ref.~\cite{Aaboud:2016hhf}, and using the LUCID-2 detector for the baseline luminosity measurements~\cite{LUCID2}.
 
The theory uncertainty of the signal yield is evaluated by varying the renormalisation and factorisation scales by factors
of $2$ and $0.5$, and using alternative PDF sets, CTEQ6L1~\cite{Pumplin:2002vw}
and MSTW2008LO~\cite{Martin:2009iq} via SysCalc~\cite{Kalogeropoulos:2018cke}.
The dominant effect on the yield comes from the scale variation.
The total effect on the signal yield is at most $10\%$. The uncertainties on the background yield are described in Section~\ref{sec:bg}.
 
\begin{table}[h]
\caption{Relative systematic uncertainties of the signal yield in the signal region, in percentage for each source.
The ranges indicate the different signal samples.
The systematic uncertainties with sub-percent contributions are not shown.}
\centerline{
\begin{tabular}{lrr}
\toprule
Component               &  Electron channel [$\%$]     &    Muon channel [$\%$]  \\
\midrule
Lepton identification      &    4--20            &   4--8   \\
Lepton isolation            &   4--5   &    1.0--1.5     \\
Lepton reconstruction   &   4--5      &  1--4       \\
Lepton trigger               &    4--5      &      0.5          \\
Pile-up                           &    < 0.5    &     2--3          \\
Luminosity                    &    2    &     2        \\
Theory                         &    10    &     10        \\
\bottomrule
\end{tabular}}
\label{tab:syst}
\end{table}

\section{Results}
\label{sec:res}
 
\begin{figure}[h]
\centering
\includegraphics[width=0.45\textwidth]{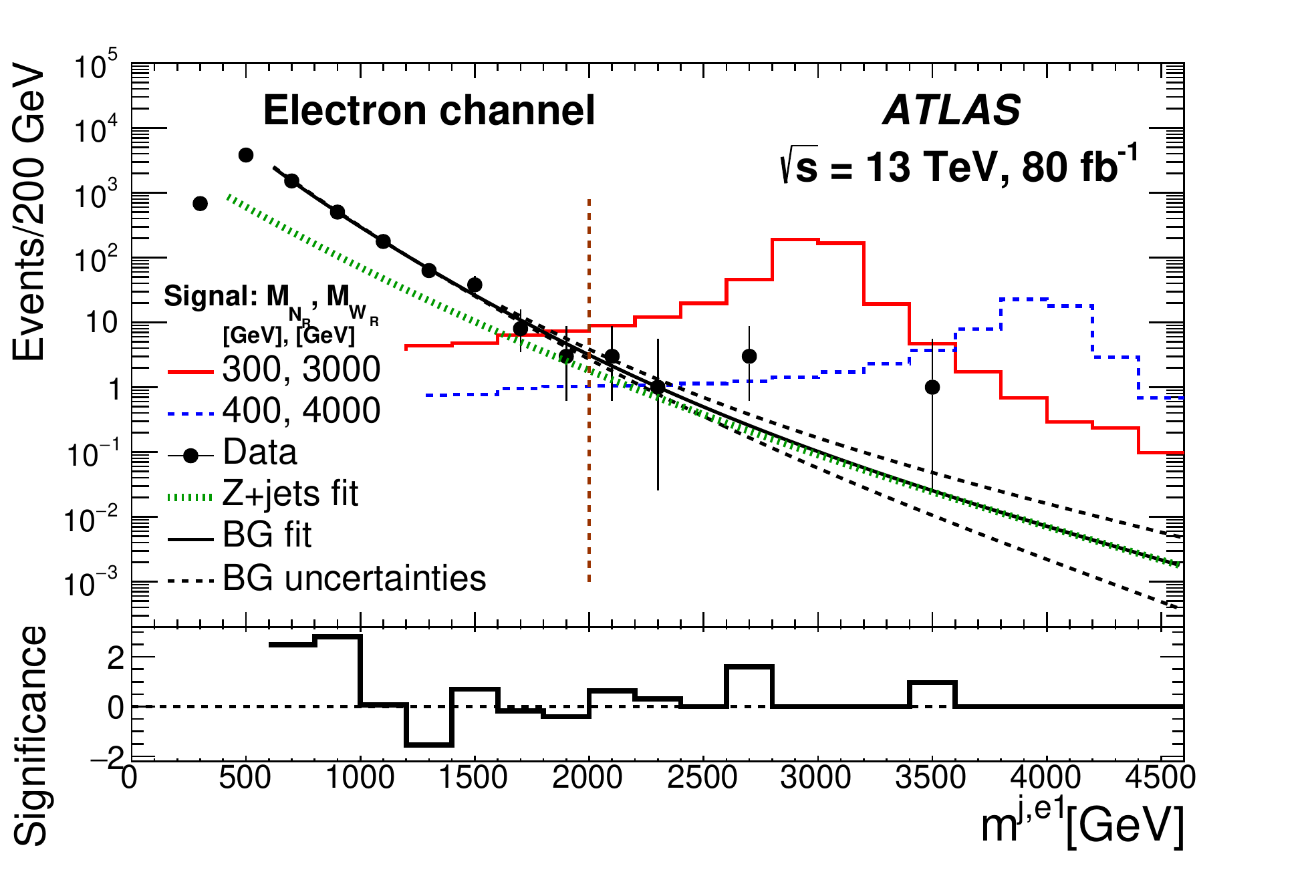}\qquad
\includegraphics[width=0.45\textwidth]{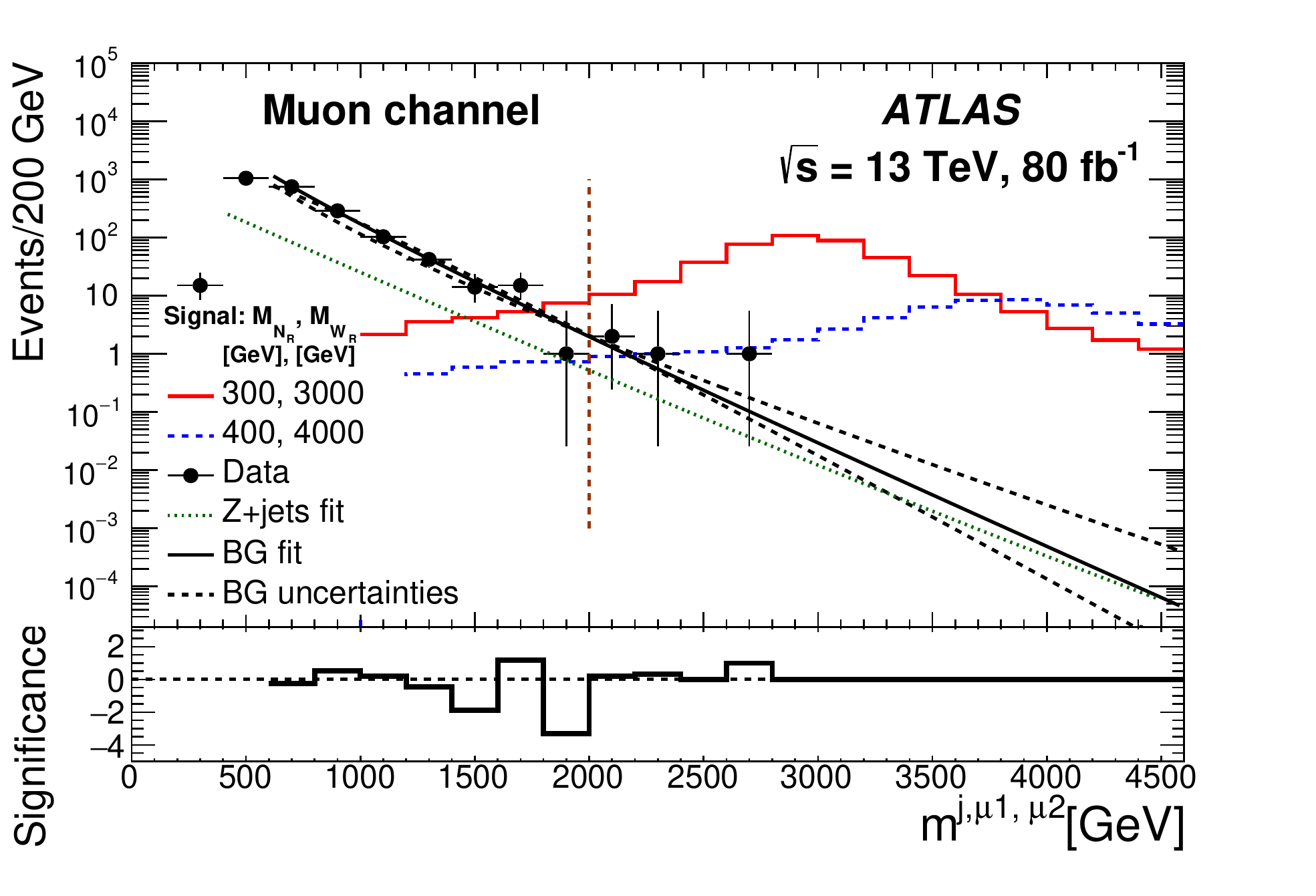}\\[1em]
\caption{Comparison of the \MWRR distribution between data and the fitted background prediction
for the electron (left) and muon (right) channels.
Two signal scenarios considered in this search are overlayed.
The dashed brown lines at $2$~\TeV\ show the boundary between the CR and SR.
The dashed black lines depict the uncertainty on the background fit.
The solid black lines on the data points indicate the Poisson uncertainties.
The significance, which indicates the deviation of data in each bin from the background fit,
is computed as the difference between the observed data and fit values, divided by the square root of the observed data value.}
\label{fig:fit-final}
\end{figure}

Figure~\ref{fig:fit-final} shows the \MWRR distributions in the SR for the electron and muon channels.
In order to search for the presence of a massive resonance, yields from simulated
signal samples and the data-driven background estimate (corresponding to $\MWRR > 2$~\TeV)
are fit to the data, separately in the electron and muon channels, using a single bin covering the entire SR.
The integral of the background functional shape in the SR is used to evaluate the expected background,
as shown in Table~\ref{tab:fsr}.
The statistical analysis is based on a likelihood fit to data.
The likelihood is constructed using a single-bin Poissonian counting-experiment approach based on the
RootStats framework~\cite{Moneta:2010pm, Baak:2014wma}.
The uncertainties of the background yield are incorporated as Gaussian constraints in the likelihood
itself in terms of a set of nuisance parameters.

The compatibility of the observed data with the background-only hypothesis is tested by
fitting the data with the background model to obtain the $p$-value.
The significance of an excess can be quantified by the probability
($p$-value) that a background-only experiment is more signal-like than observed.
The $p$-values are given in Table~\ref{tab:fsr}.
In the electron channel, 8 events are observed, while the expected background
is $2.8^{+0.5}_{-0.7}$ events.
In the muon channel 4 events are observed, while the expected background is
$1.9^{+0.5}_{-0.7}$ events.
The observed significance corresponds to 2.4 $\sigma$  in the electron channel and 1.2 $\sigma$ in the muon channel.

\begin{figure}[h]
\centering
\includegraphics[width=0.45\textwidth]{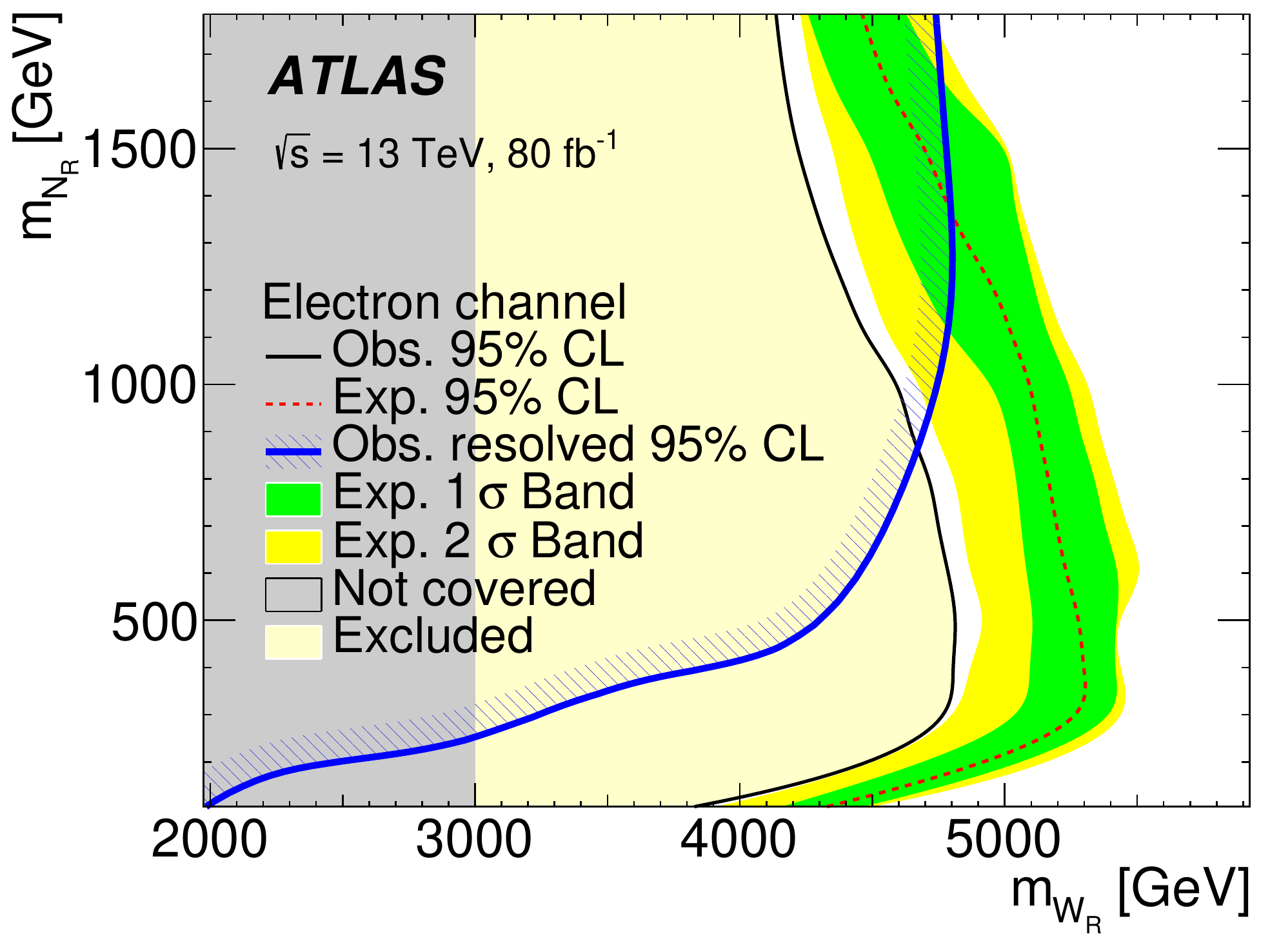}\qquad
\includegraphics[width=0.45\textwidth]{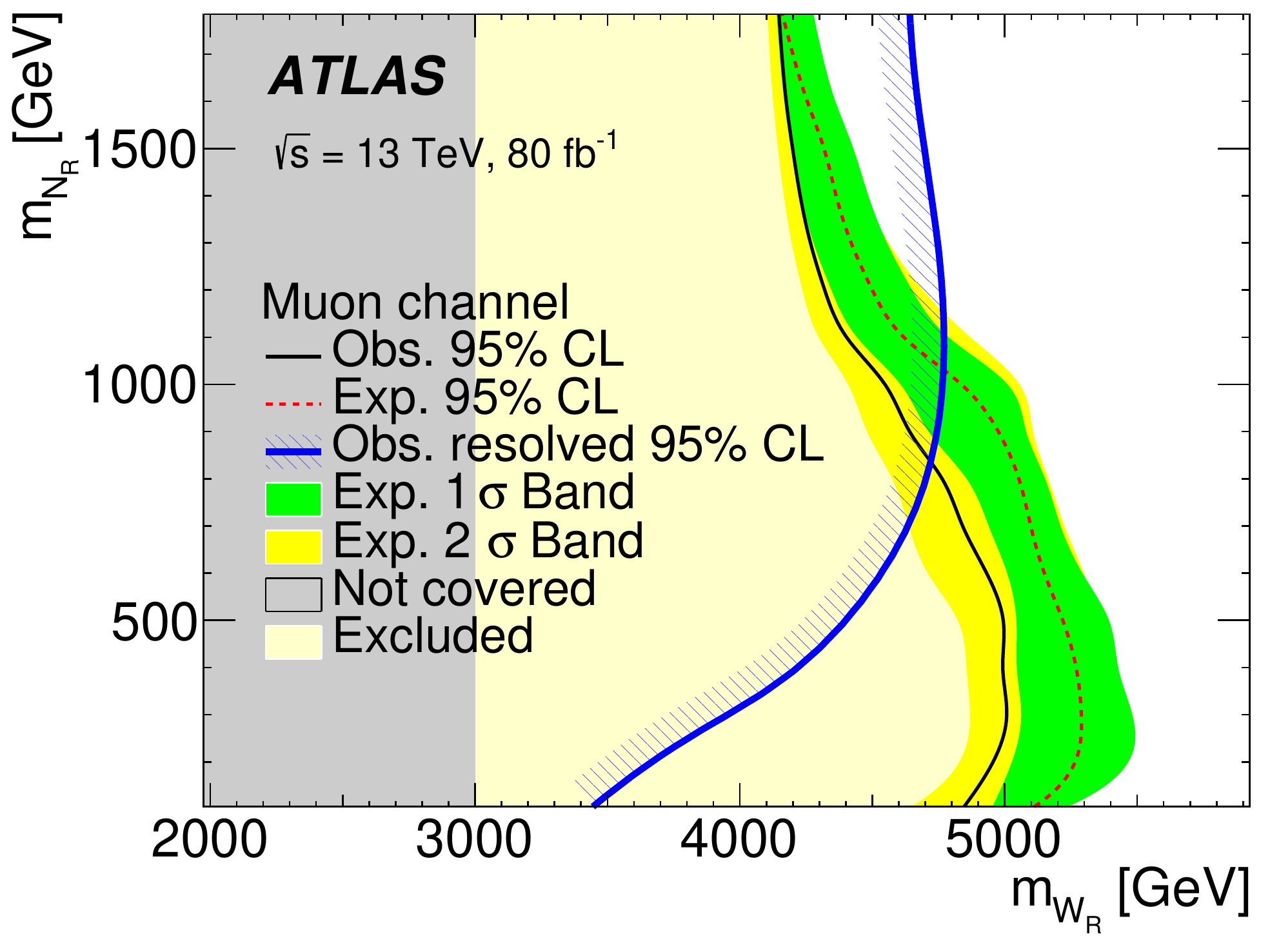}\\[1em]
\caption{
Observed (black solid line) and expected (red dashed line) 95\% CL exclusion contours in the (\MWR, \MNR) plane,
along with the $\pm 1\sigma$ and $\pm 2\sigma$ uncertainty bands (green and yellow) around the expected exclusion contour in the
electron (left) and muon (right) channels. The exclusion limits in the resolved topology~\cite{Aaboud:2018spl} are shown by the blue line.
}
\label{fig:limits}
\end{figure}

Lower limits on the masses of \NR and \WR for each of the considered signal
scenarios are determined by using the profiled likelihood
test statistic~\cite{Cowan:2010js} with the
CL$_{\textrm{s}}$ method~\cite{Junk:1999kv, Read:2002hq}.
The inputs to the limit calculations are the signal cross-sections and the
signal efficiencies in the \MWR--\MNR grid.
A linear interpolation between several benchmark samples
in the \MWR range 3--6~\TeV\ and in the \MNR range 0.1--1.8~\TeV\ is performed.
The results are shown in Figure~\ref{fig:limits}, separately for electron and muon channels.
The CL$_{\textrm{s}}$ is computed using pseudo-experiments.
The data statistical uncertainty has a significantly larger impact on the limits than the systematic uncertainties.
The leading systematic uncertainty is the background modelling uncertainty
and, in the case of the \WR and \NR mass limits, the signal theory uncertainties or the
electron identification uncertainty, depending on the signal mass values.

The excluded region extends to \MWR of 4.8~\TeV\ in the electron channel and to 5~\TeV\ in the muon channel,
for \MNR of 0.4--0.5~\TeV\, where the search is most sensitive.
For \MNR of 1.8~\TeV\,  the excluded \MWR is 4~\TeV\ in both channels.
The limits in the electron channel are weaker compared to those in the muon channel for low \MNR values,
as the electron reconstruction and identification efficiency is lower
for these \WR, \NR mass configurations.
For higher \MNR values, the worsening muon resolution and reconstruction efficiency
result in weaker limits in the muon channel.
The exclusion contour for the resolved topology~\cite{Aaboud:2018spl} is overlaid for both
channels in Figure~\ref{fig:limits},
to indicate the complementarity of the present analysis, as lower values of \MNR are excluded.

As the analysis is a single-bin Poissonian counting experiment, the limits on the cross-section are
only sensitive to the efficiencies of each signal, and do not depend significantly on \MWR and \MNR.
The observed limits on the number of selected signal events are 13.3 events
for the electron channel and 8.1 for the muon channel.
The corresponding expected limits are 5.4 events for the electron channel and 4.9 for the muon channel.

\begin{table}[h]
\caption{Observed yields and expected background yields in the signal region.
The significance and the $p$-values are shown for the background-only hypothesis.
Expected yields from three representative signal samples are also shown.}
\centerline{
\begin{tabular}{lcccc}
\toprule
&  Electron Channel & Muon Channel \\
\midrule
\addstackgap Signal ($\MWR=3~\TeV, \MNR=150~\GeV$) & $346 ^{+48}_{-75}$  & $411  ^{+36}_{-48}$  \\
\addstackgap Signal ($\MWR=3~\TeV, \MNR=300~\GeV$) & $471  ^{+42}_{-69}$  & $429  ^{+29}_{-40}$  \\
\addstackgap Signal ($\MWR=4~\TeV, \MNR=400~\GeV$) & $66 ^{+6}_{-10}$  & $57  ^{+4}_{-4} $ \\
\addstackgap Expected background & $2.8^{+0.5}_{-0.7}$ & $1.9^{+0.5}_{-0.7}$ \\
Observed events   & $8$ & $4$  \\
Significance &  2.4$\sigma$ & 1.2$\sigma$ \\
$p$-value & 0.0082 & 0.12 \\
\bottomrule
\end{tabular}}
\label{tab:fsr}
\end{table}
 
\FloatBarrier

\section{Summary}
\label{sec:summ}
 
A search for a heavy right-handed \WR boson decaying into a boosted right-handed neutrino \NR is presented using $80$~fb$^{-1}$ of $\sqrt{s} =13$~\TeV\ proton--proton collision data recorded by the ATLAS detector at the LHC.
Both electron and muon final states are analysed for the decay into two
same-flavour leptons, $\WR \to \NR \ell$, $\NR \to \ell+$jets.
In the electron final state, the analysis makes use of a \LRJ containing an electron as a proxy for \NR,
while in the muon channel, the muon four-momentum is added to the \LRJ four-momentum.
The observed \MWRR\ spectrum is consistent with the background prediction and exclusion
limits at $95\%$ confidence level are set on the \NR masses
as a function of the \WR masses.
The excluded region extends to \MWR of 4.8~\TeV\ in the electron channel and to 5~\TeV\ in the muon channel,
for \MNR of 0.4--0.5~\TeV.
The use of \LRJS results in a significant reduction of the background contribution.
Due to the signal topology, and a higher integrated luminosity this result represents an
increase of the exclusion limits in a complementary parameter space compared with previous
results that reconstruct the \WR as two resolved jets.

\section*{Acknowledgements}
 
 
We thank CERN for the very successful operation of the LHC, as well as the
support staff from our institutions without whom ATLAS could not be
operated efficiently.
 
We acknowledge the support of ANPCyT, Argentina; YerPhI, Armenia; ARC, Australia; BMWFW and FWF, Austria; ANAS, Azerbaijan; SSTC, Belarus; CNPq and FAPESP, Brazil; NSERC, NRC and CFI, Canada; CERN; CONICYT, Chile; CAS, MOST and NSFC, China; COLCIENCIAS, Colombia; MSMT CR, MPO CR and VSC CR, Czech Republic; DNRF and DNSRC, Denmark; IN2P3-CNRS, CEA-DRF/IRFU, France; SRNSFG, Georgia; BMBF, HGF, and MPG, Germany; GSRT, Greece; RGC, Hong Kong SAR, China; ISF and Benoziyo Center, Israel; INFN, Italy; MEXT and JSPS, Japan; CNRST, Morocco; NWO, Netherlands; RCN, Norway; MNiSW and NCN, Poland; FCT, Portugal; MNE/IFA, Romania; MES of Russia and NRC KI, Russian Federation; JINR; MESTD, Serbia; MSSR, Slovakia; ARRS and MIZ\v{S}, Slovenia; DST/NRF, South Africa; MINECO, Spain; SRC and Wallenberg Foundation, Sweden; SERI, SNSF and Cantons of Bern and Geneva, Switzerland; MOST, Taiwan; TAEK, Turkey; STFC, United Kingdom; DOE and NSF, United States of America. In addition, individual groups and members have received support from BCKDF, CANARIE, CRC and Compute Canada, Canada; COST, ERC, ERDF, Horizon 2020, and Marie Sk{\l}odowska-Curie Actions, European Union; Investissements d' Avenir Labex and Idex, ANR, France; DFG and AvH Foundation, Germany; Herakleitos, Thales and Aristeia programmes co-financed by EU-ESF and the Greek NSRF, Greece; BSF-NSF and GIF, Israel; CERCA Programme Generalitat de Catalunya, Spain; The Royal Society and Leverhulme Trust, United Kingdom.
 
The crucial computing support from all WLCG partners is acknowledged gratefully, in particular from CERN, the ATLAS Tier-1 facilities at TRIUMF (Canada), NDGF (Denmark, Norway, Sweden), CC-IN2P3 (France), KIT/GridKA (Germany), INFN-CNAF (Italy), NL-T1 (Netherlands), PIC (Spain), ASGC (Taiwan), RAL (UK) and BNL (USA), the Tier-2 facilities worldwide and large non-WLCG resource providers. Major contributors of computing resources are listed in Ref.~\cite{ATL-GEN-PUB-2016-002}.
 

\printbibliography

\clearpage 
 
\begin{flushleft}
{\Large The ATLAS Collaboration}

\bigskip

M.~Aaboud$^\textrm{\scriptsize 35d}$,    
G.~Aad$^\textrm{\scriptsize 101}$,    
B.~Abbott$^\textrm{\scriptsize 128}$,    
D.C.~Abbott$^\textrm{\scriptsize 102}$,    
O.~Abdinov$^\textrm{\scriptsize 13,*}$,    
A.~Abed~Abud$^\textrm{\scriptsize 70a,70b}$,    
K.~Abeling$^\textrm{\scriptsize 53}$,    
D.K.~Abhayasinghe$^\textrm{\scriptsize 93}$,    
S.H.~Abidi$^\textrm{\scriptsize 167}$,    
O.S.~AbouZeid$^\textrm{\scriptsize 40}$
% ...
\bigskip
\\

% ...
$^{13}$Institute of Physics, Azerbaijan Academy of Sciences, Baku; Azerbaijan.\\
% ...
$^{35}$$^{(a)}$Facult\'e des Sciences Ain Chock, R\'eseau Universitaire de Physique des Hautes Energies - Universit\'e Hassan II, Casablanca;$^{(b)}$Facult\'{e} des Sciences, Universit\'{e} Ibn-Tofail, K\'{e}nitra;$^{(c)}$Facult\'e des Sciences Semlalia, Universit\'e Cadi Ayyad, LPHEA-Marrakech;$^{(d)}$Facult\'e des Sciences, Universit\'e Mohamed Premier and LPTPM, Oujda;$^{(e)}$Facult\'e des sciences, Universit\'e Mohammed V, Rabat; Morocco.\\
% ...
$^{40}$Niels Bohr Institute, University of Copenhagen, Copenhagen; Denmark.\\
% ...
$^{53}$II. Physikalisches Institut, Georg-August-Universit\"{a}t G\"ottingen, G\"ottingen; Germany.\\
% ...
$^{70}$$^{(a)}$INFN Sezione di Pavia;$^{(b)}$Dipartimento di Fisica, Universit\`a di Pavia, Pavia; Italy.\\
% ...
$^{93}$Department of Physics, Royal Holloway University of London, Egham; United Kingdom.\\
% ...
$^{101}$CPPM, Aix-Marseille Universit\'e, CNRS/IN2P3, Marseille; France.\\
$^{102}$Department of Physics, University of Massachusetts, Amherst MA; United States of America.\\
% ...
$^{128}$Homer L. Dodge Department of Physics and Astronomy, University of Oklahoma, Norman OK; United States of America.\\
% ...
$^{167}$Department of Physics, University of Toronto, Toronto ON; Canada.\\
% ...
$^{a}$ Also at Centre for High Performance Computing, CSIR Campus, Rosebank, Cape Town; South Africa.\\
$^{b}$ Also at CERN, Geneva; Switzerland.\\
$^{c}$ Also at CPPM, Aix-Marseille Universit\'e, CNRS/IN2P3, Marseille; France.\\
$^{d}$ Also at D\'epartement de Physique Nucl\'eaire et Corpusculaire, Universit\'e de Gen\`eve, Gen\`eve; Switzerland.\\
$^{e}$ Also at Departament de Fisica de la Universitat Autonoma de Barcelona, Barcelona; Spain.\\
% ...
$^{*}$ Deceased

\end{flushleft}


\end{document}